\documentclass[prl,aps,showpacs,groupedaddress,longbibliography,superscriptaddress,twocolumn,toc=flat,biblatex,footinbib]{revtex4-2}
\usepackage{natbib}
\usepackage[table]{xcolor}
\usepackage[colorlinks=false]{hyperref}
\usepackage[utf8]{inputenc}
\usepackage{color}
\usepackage{bbm} 
\usepackage[english]{babel}
\usepackage{multirow}
\usepackage{amsfonts,amsmath,amssymb,stmaryrd}
\usepackage{braket}
\usepackage{graphicx}
\usepackage{subfigure}  % use for side-by-side figures
\usepackage{dsfont}
\usepackage{epsfig}
\usepackage{mathrsfs}
\usepackage{verbatim}
\usepackage{centernot}
\usepackage{ulem}
\usepackage{longtable}
\usepackage{nicefrac}
\usepackage[framemethod=tikz]{mdframed}
\usepackage{siunitx}
\usepackage[nolist]{acronym}
\usepackage{placeins}
\usepackage{soul}
\usepackage{array}
\usepackage{cancel,ifthen}
\usepackage{bm}	% bold in math mode
\renewcommand{\vec}[1]{{\bm{#1}}}
\newcommand{\mw}[1]{\textcolor{black}{#1}}
\newcommand{\mwsout}[2]{\textcolor{black}{{#1}}}

\LTcapwidth=\textwidth
\begin{document}
\normalem	

\title{Chiral Phonons {Arising From}
Chirality-Selective Magnon-Phonon Coupling
}

\author{Markus Weißenhofer}
\email[]{markus.weissenhofer@fu-berlin.de}
 \affiliation{Department of Physics and Astronomy, Uppsala University, P.\ O.\ Box 516, S-751 20 Uppsala, Sweden}
\affiliation{Department of Physics, Freie Universit{\"a}t Berlin, Arnimallee 14, D-14195 Berlin, Germany}

\author{Philipp Rieger}
\affiliation{Department of Physics and Astronomy, Uppsala University, P.\ O.\ Box 516, S-751 20 Uppsala, Sweden}
\affiliation{Department of Physics, University of Konstanz, DE-78457 Konstanz, Germany}

\author{M. S. Mrudul}
\affiliation{Department of Physics and Astronomy, Uppsala University, P.\ O.\ Box 516, S-751 20 Uppsala, Sweden}

\author{Luca Mikadze}
\affiliation{Department of Physics and Astronomy, Uppsala University, P.\ O.\ Box 516, S-751 20 Uppsala, Sweden}

\author{Ulrich Nowak}
\affiliation{Department of Physics, University of Konstanz, DE-78457 Konstanz, Germany}

\author{Peter M. Oppeneer}
\affiliation{Department of Physics and Astronomy, Uppsala University, P.\ O.\ Box 516, S-751 20 Uppsala, Sweden}

\date{\today}
%%%%%%%%%%%%%%%%%%%%%%%%%%%%%%%%%%%%%%%%%%%%%%%%%%%%%%%%%%%%%%%%%%%%%%
%%%%%%%%%%%%%%%%%%%%%%%%%%%%%%%%%%%%%%%%%%%%%%%%%%%%%%%%%%%%%%%%%%%%%%
\begin{abstract}

{Chiral phonons are desirable for applications in spintronics but their generation and control remains a challenge.}
{Here} we demonstrate the emergence of truly chiral phonons from selective magnon-phonon coupling in inversion-symmetric magnetic systems. {Considering bcc Fe as example,} we quantitatively calculate hybridized magnon-phonon quasiparticle states across the entire Brillouin zone {utilizing first-principles calculations}. Our findings challenge conventional magneto-elastic interpretations and reveal finite zero-point phonon angular momentum and strong anomalous \mw{thermal} Hall responses linked to finite (spin) Berry curvatures. {Our results further establish} that the existence of chiral phonons, particularly along high-symmetry directions, is common in many magnetic materials, offering promising avenues for novel spintronic and phononic devices.
\end{abstract}
\maketitle
\begin{acronym}
\acro{SLC}[SLC]{spin-lattice coupling}
\acro{DMI}[DMI]{Dzyaloshinskii–Moriya interaction}
\acro{dof}[DoF]{degrees of freedom}
\acro{BZ}[BZ]{Brillouin zone}
\acro{MCA}[MCA]{magneto-crystalline anisotropy}
\acro{LA}[LA]{longitudinal acoustic}
\acro{TA}[TA]{transverse acoustic}
\end{acronym}
%%%%%%%%%%%%%%%%%%%%%%%%%%%%%%%%%%%%%%%%%%%%%%%%%%%%%%%%%%%%%%%%%%%%%%
%%%%%%%%%%%%%%%%%%%%%%%%%%%%%%%%%%%%%%%%%%%%%%%%%%%%%%%%%%%%%%%%%%%%%%
\textit{Introduction.}\
Chirality, {which denotes} the asymmetry between a structure and its mirror image, is one of the fundamental properties of matter \cite{Cahn1966}. In condensed matter physics, it plays a crucial role in the properties of fermionic and bosonic (quasi-)particles, such as electrons \cite{Xu2015,Chang2018}, magnons \cite{Roessli2002,Lebrun2018,Nambu2020,Liu2022,Smejkal2023}, and phonons \cite{Zhang2015,Zhu2018,Juraschek2019,Wang2024}.

Recent theoretical predictions \cite{Zhang2015,Chen2019} and experimental observations in two-dimensional materials \cite{Zhu2018} have identified phonon modes with finite angular momentum at high-symmetry points of the \ac{BZ}, arising from the circular (or elliptical) orbital motions of atoms around their equilibrium lattice positions \cite{Zhang2014,Juraschek2019}. {Such} phonon angular momentum plays a crucial role in a wide variety of effects ranging from the phonon Hall effect \cite{Strohm2005,Grissonnanche2020,Park2020b,Flebus2023}, the ultrafast Einstein-de Haas \cite{Dornes2019,Tauchert2022} and Barnett effects \cite{Luo2023,Davies2024}, magnon-phonon conversion \cite{Holanda2018}, to
phonon magnetic moments \cite{Juraschek2017,Juraschek2019,Ren2021,Basini2024}.

While it has become common to refer to all phonons with finite angular momentum $\bm{L}$ as chiral \cite{Zhang2015,Chen2019}, symmetry arguments suggest that non-propagating phonons and those propagating in the rotation plane do not possess true chiral character \cite{Cheong2022, Ueda2023}. To unambiguously define \textit{truly} chiral phonons {\cite{Barron_2004,Juraschek2025}}, we introduce the phonon chirality parameter $\Sigma= \vec{L}\cdot \vec{v}/|\vec{v}|$, where $\vec{v}$ {is} the group velocity, based on the considerations presented in Ref.~\cite{Ueda2023}. Truly chiral phonons are characterized by a finite $\Sigma$ and have recently been experimentally detected in systems \textit{lacking} inversion ($\mathcal{P}$) symmetry {by means of X-ray and Raman spectroscopy \cite{Ueda2023,Ishito2023,Ishito2023b}, transport \cite{Kim2023,Ohe2024}, and torque measurements \cite{Zhang2025}}.

Here, we present fully quantitative, {first-principles-based} calculations to demonstrate the emergence of truly chiral phonons in the $\mathcal{P}$-symmetric ferromagnet{ic material} bcc Fe as a result of phonon-chirality-selective magnon-phonon coupling. This coupling breaks time-reversal symmetry for the phonons and is derived from a recently {developed} framework describing the interaction between lattice and magnetic \acl*{dof} \cite{Hellsvik2019,Mankovsky2022}.
We further calculate a finite zero-point angular momentum \cite{Zhang2014} and intrinsic anomalous Hall responses of the coupled magnon-phonon quasiparticles.

%%%%%%%%%%%%%%%%%%%%%%%%%%%%%%%%%%%%%%%%%%%%%%%%%%%%%%%%%%%%%%%%%%%%%%
%%%%%%%%%%%%%%%%%%%%%%%%%%%%%%%%%%%%%%%%%%%%%%%%%%%%%%%%%%%%%%%%%%%%%%
\textit{Phonon angular momentum.}\
\mwsout{Following}{As shown in} Ref.\ \cite{Zhang2014}, the angular momentum {of the phononic system} is given by
\begin{equation}
    \label{eq:L}
    \vec{L}
     = \!
    \sum_{\vec{k}\lambda} \vec{L}_{\vec{k}\lambda}\big(n_{\vec{k}\lambda}+{\textstyle \frac{1}{2}}\big)
    ,
    \hspace{0.2em}
    \mw{
    \vec{L}_{\vec{k}\lambda}
    = \!
    \sum_\mu
    2 \hbar
    \mathrm{Re}[\vec{\chi}_{\vec{k}\lambda}^{(\mu)}] \times  \mathrm{Im}[\vec{\chi}^{(\mu)}_{\vec{k}\lambda}],
    }
\end{equation}
where $n_{\vec{k}\lambda}$ is the phonon occupation number and the sum\mw{s} \mwsout{are}{is} over all phonon branches $\lambda$ and wave vectors $\vec{k}$ in the first \ac{BZ}\mw{, and over all atoms $\mu$ in the unit cell (see SM \cite{SM} for details)}. In the absence of relativistic effects, the phonon frequencies $\omega_{\vec{k}\lambda}$ and polarization vectors $\vec{\chi}_{\vec{k}\lambda}\mw{=\{\vec{\chi}_{\vec{k}\lambda}^{(\mu)}\}}$ are obtained from solving the eigenvalue problem $\mathcal{D}_\vec{k}\vec{\chi}_{\vec{k}\lambda} = \omega_{\vec{k}\lambda}^2\vec{\chi}_{\vec{k}\lambda}$ for the dynamical matrix $\mathcal{D}_\vec{k}$ \cite{Zhang2014}. In inversion symmetric systems, to which we will restrict the discussion throughout the paper, $\mathcal{D}_\vec{k}$ is real, and hence one can always choose $\vec{\chi}_{\vec{k}\lambda}=e^{i\varphi_{\vec{k}\lambda}}\vec{e}_{\vec{k}\lambda}$, with $\vec{e}_{\vec{k}\lambda}$ being real and $\varphi_{\vec{k}\lambda}$ an arbitrary phase. We call these  phonon modes \textit{linear} {(short for linearly polarized)}, as the associated dynamics of atoms are restricted to the one-dimensional space spanned by $\vec{e}_{\vec{k}\lambda}$. As such, the angular momentum $\vec{L}_{\vec{k}\lambda}$ of all linear phonon modes trivially vanishes.

At a generic nonsymmetric point of the Brillouin zone, all phonon frequencies are different and the associated eigenvectors are unique (up to a phase). Consequently, all phonon modes at such points are linear and must have zero angular momentum, irrespective of representation.
However, if two (or more) eigenvalues of the dynamical matrix coincide -- e.g.\ at high-symmetry points, lines or planes -- the eigenspace associated with this particular eigenvalue is two- (or higher-) dimensional. Considering e.g.\ two degenerate transverse phonon modes at some finite $\vec{k}$ and $\vec{v}\parallel \vec{k}$, the corresponding eigenspace can be spanned by two real and orthonormal vectors $\vec{e}_{\vec{k}1}$ and $\vec{e}_{\vec{k}2}$. They represent linear modes that can be superimposed to form left- and right-handed circular phonon modes, $\vec{\chi}_{\vec{k}1}=(\vec{e}_{\vec{k}1}+i\vec{e}_{\vec{k}2})/\sqrt{2}$ and $\vec{\chi}_{\vec{k}2}= (\vec{e}_{\vec{k}1}-i\vec{e}_{\vec{k}2})/\sqrt{2}$.
These modes correspond to the same eigenvalue as the linear modes, but they have a finite phonon chirality 
$\Sigma=\pm \hbar$, 
thus representing chiral phonon modes.

While high-symmetry regions with degenerate phonon energies are certainly interesting from a fundamental perspective, they typically have little relevance for physical effects, as they occupy regions of the {BZ} 
with virtually no volume \cite{Coh2023}. However, this changes when considering the impact of \ac{SLC} on the phonon band structure. This coupling can bridge the energy gap between two linear phonon modes, enabling them to couple and form a chiral phonon mode. Since \ac{SLC} is generally a small correction to phonon energies \cite{Ren2024}, we expect chiral phonons in inversion-symmetric systems to only emerge in regions close to degenerate points, lines, and planes of the \textit{bare} phonon spectrum \cite{note1}. This conjecture is demonstrated below, {through fully quantitative calculations} for the simple monoatomic ferromagnet bcc Fe, {which serves here as example.} The concepts can be readily applied to any other $\mathcal{P}$-symmetric magnetic system.

%%%%%%%%%%%%%%%%%%%%%%%%%%%%%%%%%%%%%%%%%%%%%%%%%%%%%%%%%%%%%%%%%%%%%%
%%%%%%%%%%%%%%%%%%%%%%%%%%%%%%%%%%%%%%%%%%%%%%%%%%%%%%%%%%%%%%%%%%%%%%
\textit{Magnon-phonon coupling.}\
To describe the coupling of spin and lattice \ac{dof} we adopt an atomistic approach. We start with the expansion of a phenomenological spin-lattice Hamiltonian up to the third order in {\ac{dof}},
\begin{align}
    \label{eq:H}
    \begin{split}
    \hat{\mathcal{H}}_\mathrm{SLC}
    &=
    \sum_i 
    \frac{\hat{\vec{P}}_i^2}{2\mw{m_i}}
    +
    \sum_{ij,\alpha\beta}
    \Big(
    \Phi_{ij}^{\alpha\beta}
    \hat{X}_i^\alpha
    \hat{X}_j^\beta
    +
    J_{ij}^{\alpha\beta}
    \hat{S}_i^\alpha
    \hat{S}_j^\beta
    \Big)
    \\
    &+
    \sum_{ijk,\alpha\beta\gamma}
    \Big(
    J_{ijk}^{\alpha\beta\gamma}
    \hat{S}_i^\alpha    
    \hat{S}_j^\beta
    \hat{X}_k^\gamma
    +
    G_{ijk}^{\alpha\beta\gamma}
    \hat{S}_i^\alpha    
    \hat{X}_j^\beta
    \hat{P}_k^\gamma
    \Big),
    \end{split}
\end{align}
keeping all terms compatible with inversion and (global) time-reversal symmetry. Here, $\hat{S}_i^\alpha$ are vector spin operators at the site $i$ with amplitude $|\vec{\hat{S}}_i|=S$ and $\hat{X}_i^\alpha$ and $\hat{P}_i^\alpha$ are nuclear displacement and momentum operators. The first three terms of $\hat{\mathcal{H}}_\mathrm{SLC}$ are the kinetic energy of the atoms (of mass $\mw{m_i}$), the lattice potential (with the force constants $\Phi_{ij}^{\alpha\beta}$ being \mw{proportional to} the Fourier transform of the dynamical matrix $\mathcal{D}_\vec{k}^{\alpha\beta}$ \cite{SM}) and the generalized Heisenberg interaction \cite{Nowak2007}, where the $J_{ij}^{\alpha\beta}$ also includes \ac{MCA}. The antisymmetric \ac{DMI} is forbidden %{
for pairs $(i,j)$ whose midpoint is an inversion center.
%\sout{by inversion symmetry}
%}

Coupling between spin and lattice \ac{dof} arises from the last two terms of $\hat{\mathcal{H}}_\mathrm{SLC}$. The term quadratic in spins and linear in displacements was recently introduced by Hellsvik \textit{et al.}~\cite{Hellsvik2019}, and since then efficient methods to calculate all $J_{ijk}^{\alpha\beta\gamma}$ parameters from first-principles have been established \cite{Mankovsky2022,Lange2023,Mankovsky2023,Miranda2024}. 

The last term 
has gained little attention in the general form expressed here. Particularly, efficient methods to calculate the full set of $G_{ijk}^{\alpha\beta\gamma}$ parameters from first principles are yet to be established. Instead, studies so far mainly investigated two special cases of this term: (i) a purely local term  $\sim \sum_i \hat{ \vec{S}}_i\cdot (\hat{ \vec{X}}_i\times \hat{ \vec{P}}_i)$ {was}
used in a semiclassical description of atomic motion 
in the effective field of the spin~\cite{Juraschek2019,Luo2023,Kahana2024}, and (ii) a term obtained by fixing the spin orientation, $\sum_{jk,\beta\gamma}G_{jk}^{\beta\gamma} \hat{X}_j^\beta\hat{P}_k^\gamma$, that appears e.g. in the Born-Huang approximation 
\cite{MeadTruhlar1979,BornHuang1996,Zhang2011,Coh2023,Ren2024}.

Magnon and phonon variables can be introduced via {the} Holstein-Primakoff transformation \cite{Holstein1940} and normal mode expansion (details in SM \cite{SM}). For a \mw{monoatomic system and expanding around a }ferromagnetic state along $z$, the magnon-phonon Hamiltonian up to second order in magnon ($\hat{b}_\vec{k}^{(\dagger)}$) and phonon operators ($\hat{a}_{\vec{k},\lambda}^{(\dagger)}$) reads
$ \hat{\mathcal{H}}_\mathrm{mp} =\sum_{\vec{k}} \hat{\mathcal{H}}_\vec{k}$ with
\begin{align}
    \label{eq:H2}
    \begin{split}
    \hat{\mathcal{H}}_\vec{k}
    &=
    \sum_{\lambda}
    \hbar 
    \omega_{\vec{k}\lambda}
    \hat{a}_{\vec{k}\lambda}^\dagger
    \hat{a}_{\vec{k}\lambda}
    +
    \varepsilon_{\vec{k}}
    \hat{b}_{\vec{k}}^\dagger
    \hat{b}_{\vec{k}}
    \\
    &+
    \sum_{\lambda}
    (
    c^-_{\vec{k}\lambda}
    \hat{b}_{-\vec{k}}
    +
    c^+_{\vec{k}\lambda}
    \hat{b}_{\vec{k}}^\dagger
    )
    (\hat{a}_{\vec{k}\lambda}
    +
    \hat{a}_{-\vec{k}\mw{\lambda}}^\dagger
    )
    \\
    &+
    \sum_{\lambda\lambda'}
    g_{\vec{k}\lambda\lambda'}
    (
    \hat{a}_{\vec{k}\lambda}
    +
    \hat{a}_{-\vec{k}\lambda}^\dagger
    )
    (
    \hat{a}_{-\vec{k}\lambda'}
    -
    \hat{a}_{\vec{k}\lambda'}^\dagger
    ).
    \end{split}
\end{align}
Here, 
$\varepsilon_{\vec{k}}$ are the bare magnon energies and zero point energies have been dropped.
The magnon-phonon couplings 
$ 
c^\pm_{\vec{k}\lambda}
=
\sum_\gamma
\sqrt{\hbar S^3/m\omega_{\vec{k}\lambda}}
( 
\tilde{J}^{xz\gamma}_{\vec{k}}
\pm
i
\tilde{J}^{yz\gamma}_{\vec{k}}
)
\chi_{\vec{k}\lambda}^\gamma$
and phonon-phonon couplings
$ g_{\vec{k}\lambda\lambda'}
=
-
\frac{i}{2}\hbar S
\sqrt{
\omega_{\vec{k}\lambda'}/\omega_{\vec{k}\lambda}
}
\sum_{\beta\gamma}
\tilde{G}^{z\beta \gamma}_{\vec{k}}
\chi^\beta_{\vec{k}\lambda}
 (\chi^\gamma_{\vec{k}\lambda'})^*$
depend on the Fourier transforms
$
\tilde{J}^{\alpha\beta\gamma}_{\vec{k}}
=
\sum_{jk} 
e^{i \vec{k}\cdot (\vec{r}_k-\vec{r}_i)}
J^{\alpha\beta\gamma}_{ijk}
$
and
$\tilde{G}^{\alpha\beta\gamma}_{\vec{k}}
=
\sum_{jk} 
e^{i \vec{k}\cdot (\vec{r}_j-\vec{r}_k)}
G^{\alpha\beta\gamma}_{ijk}$
of the {coefficients}
of the third order terms in Eq.~\eqref{eq:H}. They are relativistic corrections to the bare modes and proportional to spin-orbit coupling \cite{Mankovsky2022,Ren2024}. The emergence of chiral phonons has been {previously} 
linked to the phonon-phonon coupling term proportional to $g_{\vec{k}\lambda\lambda'}$ \cite{Zhang2011,Zhang2014,Coh2023}. Hereinafter we will omit the phonon-phonon coupling $g_{\vec{k}\lambda\lambda'}$ and instead focus on the magnon-phonon coupling $c^\pm_{\vec{k}\lambda}$. We show that magnon-phonon coupling can give rise to truly chiral phonons in bcc Fe.

For this purpose we use $J_{ijk}^{\alpha\beta\gamma}$ {coefficients}
that were recently calculated from first-principles 
\cite{Mankovsky2022}. The bare magnon energies follow $\varepsilon_{\vec{k}}=S(2d +2\sum_jJ_{ij}[ 1 - e^{-i \vec{k}\cdot (\vec{r}_j-\vec{r}_i)}])$, where the isotropic exchange constants $J_{ij}=\frac{1}{3}\sum_\alpha J^{\alpha\alpha}_{ij}$ are {also} taken from first-principles calculations \cite{Mryasov1996} and the \ac{MCA} energy 
$d$
is from experiments \cite{Razdolski2017}. The bare phonon frequencies and polarization vectors are calculated using the density-functional perturbation theory implementation of Quantum ESPRESSO \cite{Giannozzi2009,Giannozzi2017}. Both the bare magnon and phonon bandstructures are shown in the SM \cite{SM}.

%%%%%%%%%%%%%%%%%%%%%%%%%%%%%%%%%%%%%%%%%%%%%%%%%%%%%%%%%%%%%%%%%%%%%%
%%%%%%%%%%%%%%%%%%%%%%%%%%%%%%%%%%%%%%%%%%%%%%%%%%%%%%%%%%%%%%%%%%%%%%

\begin{figure}[t!]
    \centering
    \includegraphics[width=1.0\linewidth]{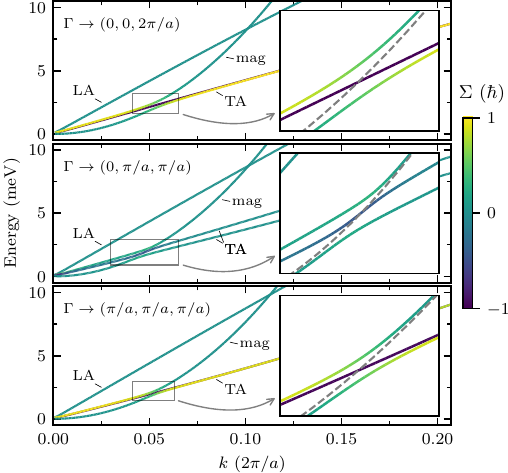}
    \caption{Coupled magnon-phonon bands in bcc Fe calculated 
    along various high-symmetry paths of the \ac{BZ} starting from  $\Gamma=(0,0,0)$.
    Labels LA, TA,  {and} mag indicate the predominant character of the mode far away from the avoided crossings and the colors encode the phonon chirality $\Sigma = \vec{L} \cdot \vec{v}/|\vec{v}|$, with $\vec{v}=\partial\omega/\partial\vec{k}$ being the group velocity. The insets zoom in on the avoided crossings (indicated by the small grey rectangles), and the grey dashed lines are the bare magnon energies. Note that the bare \ac{TA} modes are degenerate for $\Gamma \rightarrow (0,0,2\pi/a)$ and $\Gamma \rightarrow (\pi/a,\pi/a,\pi/a)$.
    }
    \label{fig:dispersion}
\end{figure}

\textit{Coupled magnon-phonon bandstructure.}\
The exact diagonalization of Eq.~\eqref{eq:H2} is performed numerically using Colpa’s method \cite{Colpa1978,SM}. The energies of the four bands along different high-symmetry paths in the \ac{BZ} are shown in Fig.~\ref{fig:dispersion}, focusing on the regions with greatest modification of the energies compared to the bare modes. The \ac{LA} phonon does not couple to magnons and is thus unaltered by \ac{SLC}. 

We observe \textit{avoided crossings} where the bare modes intersect, which indicate the formation of hybrid magnon-phonon quasiparticles, the so-called \textit{magnon polarons} \cite{Li2021}. The energy gaps between the hybridizing modes range from around $\SI{0.18}{\milli\electronvolt}$ to $\SI{0.45}{\milli\electronvolt}$. Physically, the magnon-phonon coupling contains an antisymmetric, \ac{DMI}-like contribution,
$ 
c^{\pm,a}_{\vec{k}\lambda}
=
\frac{1}{2}
\sum_\gamma
\sqrt{\hbar S^3/m\omega_{\vec{k}\lambda}}
[
( 
\tilde{J}^{xz\gamma}_{\vec{k}}
-
\tilde{J}^{zx\gamma}_{\vec{k}}
)
\pm
i
(
\tilde{J}^{yz\gamma}_{\vec{k}}
-
\tilde{J}^{zy\gamma}_{\vec{k}}
)
]
\chi_{\vec{k}\lambda}^\gamma$ 
and a symmetric, two-site anisotropy-like contribution
$ 
c^{\pm,s}_{\vec{k}\lambda}
=
\frac{1}{2}
\sum_\gamma
\sqrt{\hbar S^3/m\omega_{\vec{k}\lambda}}
[
( 
\tilde{J}^{xz\gamma}_{\vec{k}}
+
\tilde{J}^{zx\gamma}_{\vec{k}}
)
\pm
i
(
\tilde{J}^{yz\gamma}_{\vec{k}}
+
\tilde{J}^{zy\gamma}_{\vec{k}}
)
]
\chi_{\vec{k}\lambda}^\gamma$,
both of which are of relativistic origin. In bcc Fe, the \ac{DMI}-like coupling greatly exceeds the symmetric one \cite{Mankovsky2022}. {Therefore,} the widely-used conventional magneto-elastic theory \cite{Kittel1949} is unable to describe the magnon-phonon dispersion calculated here, as it lacks a term related to \ac{DMI}. Such a term {was} only recently derived \cite{Weissenhofer2023} from the atomistic \ac{SLC} Hamiltonian \eqref{eq:H}. In {the} light of this so-far overlooked \ac{DMI}-like coupling, established interpretations of magnon-phonon hybridization based on conventional magneto-elastic theory \cite{White1965,Rueckriegel2014,Kamra2015,Streib2019,Gurevich2020} 
have to be reconsidered.

As 
%a 
{one}
key result of this paper, we find that the magnon mode only hybridizes with one of the \ac{TA} modes, if the two bare \ac{TA} phonons are degenerate. In the non-degenerate case, i.e., $\Gamma$ to $(0,\pi/a,\pi/a)$, there are two avoided crossings with both \ac{TA} modes in close proximity. Calculating the phonon chirality $\Sigma$ {-- which is obtained using the angular momentum eigenvalues of the hybridized system rather than the ones of the bare phonons \cite{SM} --} for these magnon-phonon modes, we reveal that the magnons selectively couple to phonons with one chirality. We note that upon reversal of the magnetization from $z$ to $-z$, the magnons instead only couple to phonons with opposite chirality. We {further} emphasize that the magnon-phonon coupling also lifts the degeneracy of the phonons far away from the avoided crossing regions, leading to a small energy gap of the order of $\SI{1}{\micro\electronvolt}$ between the two chiral phonon modes. {
This value is comparable to that predicted from phonon-phonon coupling \cite{Coh2023}. However, it remains to be determined whether such a small gap persists when the spectrum is broadened and renormalized by higher-order phonon-phonon or magnon-phonon scattering.}

\mw{Qualitatively, the selective coupling of magnons to chiral phonons with positive $\Sigma$ can be understood by noting that both spin precession and atomic revolution occur in the same counterclockwise direction \cite{note3}. A more quantitative way to understand this phenomenon is to first express the bare phonons in a circular basis before calculating their hybridization with magnons. This approach is only valid along high-symmetry paths in the BZ where the bare TA phonon modes are degenerate (e.g., the paths shown in the top and bottom panels of Fig.~\ref{fig:dispersion}). Notably, in this basis, the magnon-phonon coupling $c^\pm_{\vec{k}\lambda}$ for one of the phonon modes, specifically that of circular phonons with $\Sigma = -\hbar$, vanishes exactly.}
Selective hybridization between magnons and non-propagating circularly polarized phonons has been measured recently in the layered zigzag antiferromagnet FePSe$_3$ \cite{Cui2023}.

\begin{figure}[t]
    \centering
    \includegraphics[width=1.0\linewidth]{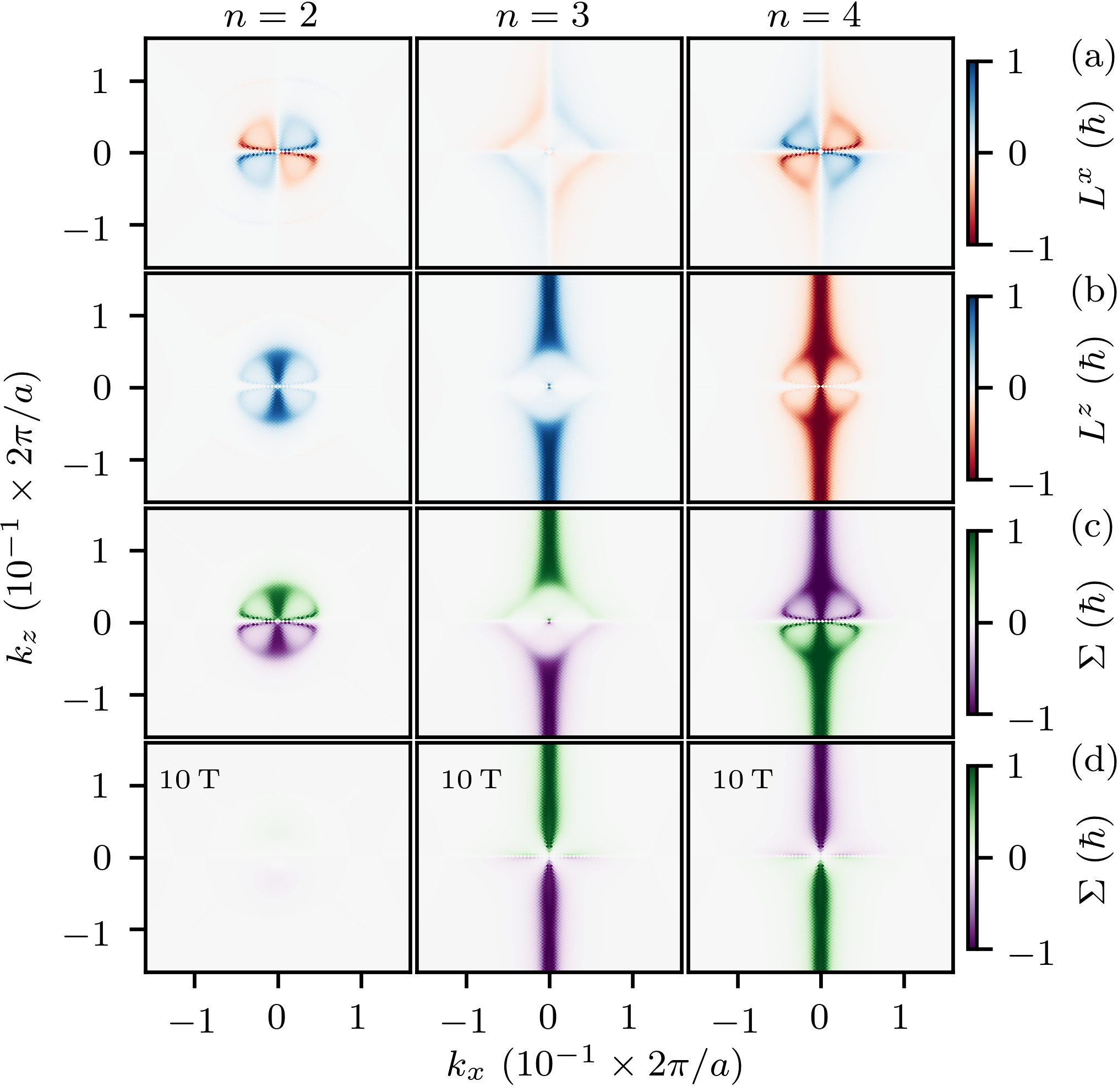}
    \caption{Phonon angular momenta and chiralities of the coupled magnon-phonon bands {computed} for bcc Fe. Plots in each row illustrate (a) $L^x$, (b) $L^z$, (c) $\Sigma$ without applied magnetic field, and (d) $\Sigma$ with a field of $B=\SI{10}{\tesla}$ for the different magnon-phonon modes with label $n\in \{2,3,4\}$. The mode with $n=1$ (\ac{LA} phonon) is achiral and hence not shown. $L^y$ is zero in the $k_x$-$k_z$ plane depicted here.}
    \label{fig:angMom}
\end{figure}

The phonon angular momentum and chirality in the $k_x$-$k_z$ plane of the \ac{BZ} are shown in Fig.\ \ref{fig:angMom} (results for other planes are discussed in the SM \cite{SM}).
As argued above, truly chiral phonons primarily occur in close proximity of the high symmetry axes -- here, the $k_z$ axis -- where they arise as a superposition of the degenerate bare \ac{TA} phonons. Note that phonons along $k_x$ and $k_y$ remain linear. This symmetry breaking arises from the orientation of magnetization along the $z$-direction in combination with spin-orbit coupling \cite{Dresselhaus2007}. 

The emergence of chiral phonons in a roughly circular pattern around the $\Gamma$ point {[see Fig.\ \ref{fig:angMom}(a),(b)]} is a result of magnon-phonon hybridization. This can be demonstrated by applying an external magnetic field in the direction of the magnetization. A magnetic field $B$ shifts the bare magnon energies by $\varepsilon_{\vec{k}}\rightarrow \varepsilon_{\vec{k}}+\mu_\mathrm{s}B$, with $\mu_\mathrm{s}=2.2\,\mu_\mathrm{B}$ being the saturation magnetic moment of bcc Fe \cite{Evans2014}. 
For field strengths of $B = \SI{10}{\tesla}$ the energies of the bare magnons are way above those of the \ac{TA} phonons \cite{note2}, making avoided crossings impossible and confining the chiral phonons to the vicinity of the $k_z$-
direction, {see Fig.\ \ref{fig:angMom}(d)}. 
{This proves} that magnon-phonon coupling can induce chiral phonons without {direct} hybridization with magnons, i.e., if their energies are very different.

\begin{figure}
    \centering
    \includegraphics[width=1.0\linewidth]{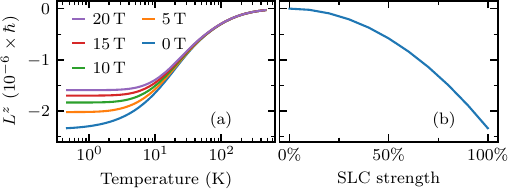}
    \caption{Nonzero component {$L_z$} of the {equilibrium} phonon angular momentum per unit cell for bcc Fe, {shown in} (a) {versus} temperature and for applied magnetic fields as labeled and {in} (b) {versus} rescaled \ac{SLC} strength.}
    \label{fig:totAngMom}
\end{figure}

Next, we calculate the equilibrium phonon angular momentum for bcc Fe via Eq.~\eqref{eq:L}, see Fig.~\ref{fig:totAngMom}. In thermal equilibrium the occupation numbers of a magnon-phonon state with energy $\varepsilon_{\vec{k}n}$ follow the Bose-Einstein distribution, $n_{\vec{k}n}=1/(e^{\varepsilon_{\vec{k}n}/k_\mathrm{B}T}-1)$. The scaling of the {equilibrium} phonon angular momentum with temperature $T$ is shown in Fig.~\ref{fig:totAngMom}(a). \mwsout{W}{Surprisingly, w}e find that the system has a small, but finite \textit{zero-point angular momentum} $ \vec{L}(T\rightarrow0)=    \sum_{\vec{k}n} \frac{1}{2} \vec{L}_{\vec{k}n}$ \cite{Zhang2014}, because the angular momenta of the different modes do not fully cancel. {This is a result of the nonunitary nature of the Bogoliubov-Valatin transformation applied within Colpa's method \cite{SM}}. Moreover, it is observed that the {equilibrium} phonon angular momentum vanishes in the limit of high temperatures.
{
It can be proven analytically that $\vec{L}(T\rightarrow \infty)=0$ is true for all inversion-symmetric systems with magnon-phonon coupling~\cite{SM}.
%\sout{This is a completely general feature of phonons in inversion-symmetric systems, which can be proven analytically as follows. In general, the thermal average of any observable $\hat{O}$ is given by $\langle\hat{O}\rangle=\mathrm{Tr}[e^{-\hat{\mathcal{H}}/k_\mathrm{B}T}\hat{O}]/\mathrm{Tr}[e^{-\hat{\mathcal{H}}/k_\mathrm{B}T}]$, which for $T\rightarrow\infty$ simplifies to $\langle\hat{O}\rangle=\mathrm{Tr}[\hat{O}]/d$, $d$ being the dimension of the Hilbert space. Since the trace is basis independent, we can conveniently choose the bare phonon and magnon modes as basis. In an inversion-symmetric system, we can in particular choose linear phonon modes, in which case the angular momentum operator $\hat{\vec{L}}$ has vanishing diagonal matrix elements. It follows immediately that the thermal expectation value of the phonon angular momentum vanishes for $T\rightarrow \infty$.
%Note that our proof is more general than the one presented in Ref.~\cite{Zhang2014}, which only proves $L^z(T\rightarrow\infty)=0$ for two-dimensional phonons subject to a Lorentz force.
%}
}

By changing the applied magnetic field we can eliminate the impact of \ac{BZ} regions with avoided crossings on $\vec{L}$, revealing the dominant role of chiral phonons without substantial hybridization with magnons in the high-temperature limit. To further elucidate the concept of zero-point angular momentum, we vary the strength of \ac{SLC} from $0\%$ to $100\%$ [Fig.~\ref{fig:totAngMom}(b)]. This scaling highlights that the angular momentum arises solely from relativistic effects, emphasizing the intrinsic link between spin-orbit coupling and the observed zero-point contributions.

%%%%%%%%%%%%%%%%%%%%%%%%%%%%%%%%%%%%%%%%%%%%%%%%%%%%%%%%%%%%%%%%%%%%%%
%%%%%%%%%%%%%%%%%%%%%%%%%%%%%%%%%%%%%%%%%%%%%%%%%%%%%%%%%%%%%%%%%%%%%%
\textit{Intrinsic anomalous Hall responses.}\
Nontrivial (spin) Berry curvature arising due to magnon-polaron formation can lead to the emergence of Hall-type quantum transport phenomena of heat and spin angular momentum, manifested in thermal Hall and spin Nernst effects \cite{Park2020,Klogetvedt2023,Bao2023}.
The Berry curvature of the $n$th band can be obtained by casting the Hamiltonian in a Bogoliubov-de Gennes form \cite{Mook2019}, 
{which gives}
\begin{align}
    \Omega_{\mu\nu}^n(\vec{k})
    &=
    2
    i
    \hbar^2
    \sum_{m\neq n}^{2N_\mathrm{b}}
    g_{nn}
    g_{mm}
    \frac{
    \bra{n_{\vec{k}}}
    \hat{v}_\mu
    \ket{m_{\vec{k}}}
    \bra{m_{\vec{k}}}
    \hat{v}_\nu
    \ket{n_{\vec{k}}}
    }{[(g\mathcal{E}_\vec{k})_{nn}-(g\mathcal{E}_\vec{k})_{mm}]^2},
\end{align}
where $\ket{n_{\vec{k}}}$ is an eigenstate of $\hat{\mathcal{H}}_{\vec{k}}$, $N_\mathrm{b}$ is the number of magnon-phonon bands, $\mathcal{E}_\vec{k}=\mathrm{diag}(\varepsilon_{\vec{k}1},\dots, \varepsilon_{\vec{k}N_\mathrm{b}},$ $\varepsilon_{-\vec{k}1},\dots,\varepsilon_{-\vec{k}N_\mathrm{b}}  )$ is a matrix containing the eigenenergies,  $\hat{\vec{v}}= \hbar^{-1} \partial_\vec{k}\hat{\mathcal{H}}_{\vec{k}}$ is the velocity operator and $g=\sigma_z \otimes \mathds{1}_{N_\mathrm{d} \times N_\mathrm{d}}$, with $\sigma_z$ being the $z$ component of the Pauli matrices and $\mathds{1}_{N_\mathrm{b} \times N_\mathrm{b}}$ the unit matrix of dimension $N_\mathrm{b}$. For details, we refer to Ref.~\cite{Klogetvedt2023}.

Hereinafter, we consider an ultrathin, quasi-two dimensional layer of bcc Fe by restricting the $\vec{k}$ values to a plane through the \ac{BZ} that includes $\Gamma$.  Integrating the Berry curvature of the $n$th band over such plane yields the respective first Chern number $C_n$. E.g., for the $k_x$-$k_y$ plane it reads $C_n=\frac{1}{2\pi}\int dk_x dk_y \Omega_{xy}^n(\vec{k})$. Even though we obtain finite Berry curvatures as a result of time-reversal symmetry breaking via the \ac{SLC}, Chern numbers of the magnon-phonon bands are zero, i.e., the bandstructure is topologically trivial \cite{Chang2023}. This is because there are two topological gaps, one at the aforementioned avoided crossing region and a second one close to $\Gamma$, with an opposite sign of Berry curvature. Note that for calculating the Chern numbers, we apply a small magnetic field of  $B\geqslant \SI{0.5}{\tesla}$ to broaden the Berry curvature close to $\Gamma$ \mw{and converge the computations with reasonable numerical effort}.

The anomalous {magnon-phonon} thermal Hall conductivity $\kappa_{\mu\nu}$ and the spin Nernst coefficient $\alpha^{S_\tau}_{\mu\nu}$, respectively, relate a transverse heat current and a transverse spin current to an applied temperature gradient, expressed as $ J_{\mu}=-\sum_\nu\kappa_{\mu\nu}\partial_\nu T$ and $J^{S_\tau}_{\mu}=-\sum_\nu\alpha^{S_\tau}_{\mu\nu}\partial_\nu T$. Within linear response theory, $\kappa_{\mu\nu}$ is related to the Berry curvature via \cite{Zhang2016,Matsumoto2014,Murakami2017,Zhang2024}
\begin{align}
    \kappa_{\mu\nu}
    &=
    -\frac{k_\mathrm{B}^2 T}{\hbar \mathcal{A}}
    \sum_{\vec{k}}
    \sum_{n=1}^{N_\mathrm{b}}
    c_2(n_{\vec{k}n})
    \Omega_{\mu\nu}^n(\vec{k}),
\end{align}
where $\mathcal{A}$ is the area of the system and $c_2(x)= (1+x)(\ln \frac{1+x}{x})^2-(\ln x)^2 - 2 \mathrm{Li}_2(-x)$, with $\mathrm{Li}_2(x)$ being the second order polylogarithm function.
Similarly, the spin Nernst coefficient $\alpha^{S_\tau}_{\mu\nu}$ can be obtained from the spin Berry curvature $\Omega_{\mu\nu}^{S_\tau,n}(\vec{k})$ via \cite{Go2022,Park2020,Li2020}
\begin{align}
    \alpha^{S_\tau}_{\mu\nu}
    &=
    -\frac{2k_\mathrm{B} }{\mathcal{A}}
    \sum_{\vec{k}}
    \sum_{n=1}^{N_\mathrm{b}}
    c_1(\mw{n_{\vec{k}n}})
    \Omega_{\mu\nu}^{S_\tau,n}(\vec{k}), \\
    \Omega_{\mu\nu}^{S_\tau,n}(\vec{k})
    &=
    2
    i
    \hbar^2
    \sum_{m\neq n}^{2N_\mathrm{b}}
    g_{nn}
    g_{mm}
    \frac{
    \bra{n_{\vec{k}}}
    \hat{j}_\mu^{S_\tau}
    \ket{m_{\vec{k}}}
    \bra{m_{\vec{k}}}
    \hat{v}_\nu
    \ket{n_{\vec{k}}}
    }{[(g\mathcal{E}_\vec{k})_{nn}-(g\mathcal{E}_\vec{k})_{mm}]^2},
\end{align}
with $c_1(x)=(1+x)\ln(1+x)-x\ln x$, and $\hat{j}_\mu^{S_\tau}=\frac{1}{4}\{\hat{v}_\mu,g\hat{S}_\tau\}$.

\begin{figure}[t]
    \centering
    \includegraphics[width=1.0\linewidth]{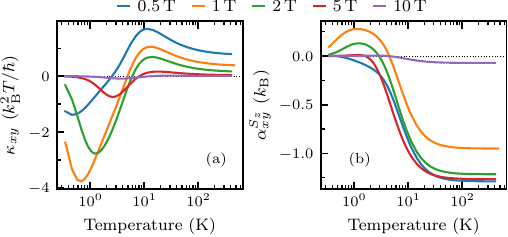}
    \caption{Anomalous thermal Hall conductivity (a) and spin Nernst coefficient (b) {due to} magnon-polarons in an ultrathin Fe(001) film. The magnetization is in out-of-plane direction and a magnetic field with values as labeled is applied in the same direction.}
    \label{fig:transCond}
\end{figure}

Both anomalous transport coefficients, calculated for an ultrathin Fe(001) film by summing $\vec{k}$ over the $k_x$-$k_y$ plane, are shown in Fig.\ \ref{fig:transCond}. They turn out to be significantly larger than what has been calculated {previously} for magnon-polaron bands in honeycomb ferro- \cite{Klogetvedt2023} and ferrimagnets \cite{Park2020}. This is because in bcc Fe the magnon-phonon modes with finite (spin) Berry curvature have rather low energies, thus contributing strongly to transport. By varying the applied magnetic field, we {further} demonstrate the tunability of $\kappa_{xy}$ and $\alpha^{S_z}_{xy}$. As mentioned earlier, strong applied fields suppress the hybridization of magnons and phonons by shifting the bare magnon energies above those of the transverse acoustic \ac{TA} phonons. This underscores that the emergence of intrinsic anomalous Hall responses is fundamentally linked to the formation of magnon-polaron hybrid bands.

%%%%%%%%%%%%%%%%%%%%%%%%%%%%%%%%%%%%%%%%%%%%%%%%%%%%%%%%%%%%%%%%%%%%%%
%%%%%%%%%%%%%%%%%%%%%%%%%%%%%%%%%%%%%%%%%%%%%%%%%%%%%%%%%%%%%%%%%%%%%%
\textit{Conclusions.}\
We have demonstrated the existence of truly chiral phonons arising from chirality-selective magnon-phonon coupling in inversion-symmetric magnetic systems. Our {first-principles-based}  {approach provides a robust, quantitative framework for understanding the hybridized magnon-phonon quasiparticle states across the entire BZ.} Given that magnon-phonon coupling in bcc Fe is primarily influenced by a DMI-like term, we conclude that 
{previous} interpretations of magnon-phonon hybridization based on conventional magneto-elastic theory, which does not account for such a term, must be reconsidered. 
{We further reveal the existence of a finite zero-point phonon angular momentum -- an entirely quantum mechanical phenomenon without classical counterpart -- emerging from the intricate coupling between phononic zero-point fluctuations and those of the spin angular momenta. }
Additionally, we observe strong and tunable anomalous Hall responses arising from finite (spin) Berry curvatures associated with magnon-phonon hybridization.

{Our results imply that}
the existence of truly chiral phonons along high-symmetry directions, featuring at least two degenerate bare phonon modes, 
{is} an abundant characteristic in many magnetic materials. 
{Since the ability to control and utilize these chiral phonons opens new avenues for transporting angular momentum and manipulating magnetic order}, this finding will certainly prove useful in the search for potential material candidates for novel spintronic or phononic devices.

%%%%%%%%%%%%%%%%%%%%%%%%%%%%%%%%%%%%%%%%%%%%%%%%%%%%%%%%%%%%%%%%%%%%%%
%%%%%%%%%%%%%%%%%%%%%%%%%%%%%%%%%%%%%%%%%%%%%%%%%%%%%%%%%%%%%%%%%%%%%%
%\acknowledgements
{
M.W.\ and P.M.O.\ acknowledge
 support from the German Research Foundation (Deutsche Forschungsgemeinschaft) through CRC/TRR 227 ``Ultrafast Spin Dynamics" (Project
MF, Project ID No.\ 328545488).
\mw{M.W.\ and M.S.M.\ thank José Ángel Castellanos-Reyes for stimulating discussions.}
U.N.\ acknowledges support from the Germany Research Foundation through CRC 1432 \mw{(project no. 425217212) and the RU ChiPS (project no. 541503763)}.
This work has furthermore been  supported by the Swedish Research Council (VR), the Knut and Alice Wallenberg Foundation (Grants No.\ 2022.0079 and No.\ 2023.0336), {and by the EIC Pathfinder OPEN grant No.\ 101129641 (OBELIX).} The calculations were enabled by resources provided by the National Academic Infrastructure for Supercomputing in Sweden (NAISS) at NSC Link\"oping partially funded by the Swedish Research Council through grant agreement No.\ 2022-06725.
}
%%%%%%%%%%%%%%%%%%%%%%%%%%%%%%%%%%%%%%%%%%%%%%%%%%%%%%%%%%%%%%%%%%%%%%
%%%%%%%%%%%%%%%%%%%%%%%%%%%%%%%%%%%%%%%%%%%%%%%%%%%%%%%%%%%%%%%%%%%%%%
%\bibliography{bibfile.bib}

\begin{thebibliography}{79}%
\makeatletter
\providecommand \@ifxundefined [1]{%
 \@ifx{#1\undefined}
}%
\providecommand \@ifnum [1]{%
 \ifnum #1\expandafter \@firstoftwo
 \else \expandafter \@secondoftwo
 \fi
}%
\providecommand \@ifx [1]{%
 \ifx #1\expandafter \@firstoftwo
 \else \expandafter \@secondoftwo
 \fi
}%
\providecommand \natexlab [1]{#1}%
\providecommand \enquote  [1]{``#1''}%
\providecommand \bibnamefont  [1]{#1}%
\providecommand \bibfnamefont [1]{#1}%
\providecommand \citenamefont [1]{#1}%
\providecommand \href@noop [0]{\@secondoftwo}%
\providecommand \href [0]{\begingroup \@sanitize@url \@href}%
\providecommand \@href[1]{\@@startlink{#1}\@@href}%
\providecommand \@@href[1]{\endgroup#1\@@endlink}%
\providecommand \@sanitize@url [0]{\catcode `\\12\catcode `\$12\catcode
  `\&12\catcode `\#12\catcode `\^12\catcode `\_12\catcode `\%12\relax}%
\providecommand \@@startlink[1]{}%
\providecommand \@@endlink[0]{}%
\providecommand \url  [0]{\begingroup\@sanitize@url \@url }%
\providecommand \@url [1]{\endgroup\@href {#1}{\urlprefix }}%
\providecommand \urlprefix  [0]{URL }%
\providecommand \Eprint [0]{\href }%
\providecommand \doibase [0]{https://doi.org/}%
\providecommand \selectlanguage [0]{\@gobble}%
\providecommand \bibinfo  [0]{\@secondoftwo}%
\providecommand \bibfield  [0]{\@secondoftwo}%
\providecommand \translation [1]{[#1]}%
\providecommand \BibitemOpen [0]{}%
\providecommand \bibitemStop [0]{}%
\providecommand \bibitemNoStop [0]{.\EOS\space}%
\providecommand \EOS [0]{\spacefactor3000\relax}%
\providecommand \BibitemShut  [1]{\csname bibitem#1\endcsname}%
\let\auto@bib@innerbib\@empty
%</preamble>
\bibitem [{\citenamefont {Cahn}\ \emph {et~al.}(1966)\citenamefont {Cahn},
  \citenamefont {Ingold},\ and\ \citenamefont {Prelog}}]{Cahn1966}%
  \BibitemOpen
  \bibfield  {author} {\bibinfo {author} {\bibfnamefont {R.~S.}\ \bibnamefont
  {Cahn}}, \bibinfo {author} {\bibfnamefont {C.}~\bibnamefont {Ingold}},\ and\
  \bibinfo {author} {\bibfnamefont {V.}~\bibnamefont {Prelog}},\ }\bibfield
  {title} {\bibinfo {title} {{Specification of Molecular Chirality}},\ }\href
  {https://doi.org/https://doi.org/10.1002/anie.196603851} {\bibfield
  {journal} {\bibinfo  {journal} {Angew. Chem. Intern. Ed.}\ }\textbf {\bibinfo
  {volume} {5}},\ \bibinfo {pages} {385} (\bibinfo {year} {1966})}\BibitemShut
  {NoStop}%
\bibitem [{\citenamefont {Xu}\ \emph {et~al.}(2015)\citenamefont {Xu},
  \citenamefont {Belopolski}, \citenamefont {Alidoust}, \citenamefont
  {Neupane}, \citenamefont {Bian}, \citenamefont {Zhang}, \citenamefont
  {Sankar}, \citenamefont {Chang}, \citenamefont {Yuan}, \citenamefont {Lee},
  \citenamefont {Huang}, \citenamefont {Zheng}, \citenamefont {Ma},
  \citenamefont {Sanchez}, \citenamefont {Wang}, \citenamefont {Bansil},
  \citenamefont {Chou}, \citenamefont {Shibayev}, \citenamefont {Lin},
  \citenamefont {Jia},\ and\ \citenamefont {Hasan}}]{Xu2015}%
  \BibitemOpen
  \bibfield  {author} {\bibinfo {author} {\bibfnamefont {S.-Y.}\ \bibnamefont
  {Xu}}, \bibinfo {author} {\bibfnamefont {I.}~\bibnamefont {Belopolski}},
  \bibinfo {author} {\bibfnamefont {N.}~\bibnamefont {Alidoust}}, \bibinfo
  {author} {\bibfnamefont {M.}~\bibnamefont {Neupane}}, \bibinfo {author}
  {\bibfnamefont {G.}~\bibnamefont {Bian}}, \bibinfo {author} {\bibfnamefont
  {C.}~\bibnamefont {Zhang}}, \bibinfo {author} {\bibfnamefont
  {R.}~\bibnamefont {Sankar}}, \bibinfo {author} {\bibfnamefont
  {G.}~\bibnamefont {Chang}}, \bibinfo {author} {\bibfnamefont
  {Z.}~\bibnamefont {Yuan}}, \bibinfo {author} {\bibfnamefont {C.-C.}\
  \bibnamefont {Lee}}, \bibinfo {author} {\bibfnamefont {S.-M.}\ \bibnamefont
  {Huang}}, \bibinfo {author} {\bibfnamefont {H.}~\bibnamefont {Zheng}},
  \bibinfo {author} {\bibfnamefont {J.}~\bibnamefont {Ma}}, \bibinfo {author}
  {\bibfnamefont {D.~S.}\ \bibnamefont {Sanchez}}, \bibinfo {author}
  {\bibfnamefont {B.}~\bibnamefont {Wang}}, \bibinfo {author} {\bibfnamefont
  {A.}~\bibnamefont {Bansil}}, \bibinfo {author} {\bibfnamefont
  {F.}~\bibnamefont {Chou}}, \bibinfo {author} {\bibfnamefont {P.~P.}\
  \bibnamefont {Shibayev}}, \bibinfo {author} {\bibfnamefont {H.}~\bibnamefont
  {Lin}}, \bibinfo {author} {\bibfnamefont {S.}~\bibnamefont {Jia}},\ and\
  \bibinfo {author} {\bibfnamefont {M.~Z.}\ \bibnamefont {Hasan}},\ }\bibfield
  {title} {\bibinfo {title} {{Discovery of a Weyl fermion semimetal and
  topological Fermi arcs}},\ }\href {https://doi.org/10.1126/science.aaa9297}
  {\bibfield  {journal} {\bibinfo  {journal} {Science}\ }\textbf {\bibinfo
  {volume} {349}},\ \bibinfo {pages} {613} (\bibinfo {year}
  {2015})}\BibitemShut {NoStop}%
\bibitem [{\citenamefont {Chang}\ \emph {et~al.}(2018)\citenamefont {Chang},
  \citenamefont {Wieder}, \citenamefont {Schindler}, \citenamefont {Sanchez},
  \citenamefont {Belopolski}, \citenamefont {Huang}, \citenamefont {Singh},
  \citenamefont {Wu}, \citenamefont {Chang}, \citenamefont {Neupert},
  \citenamefont {Xu}, \citenamefont {Lin},\ and\ \citenamefont
  {Hasan}}]{Chang2018}%
  \BibitemOpen
  \bibfield  {author} {\bibinfo {author} {\bibfnamefont {G.}~\bibnamefont
  {Chang}}, \bibinfo {author} {\bibfnamefont {B.~J.}\ \bibnamefont {Wieder}},
  \bibinfo {author} {\bibfnamefont {F.}~\bibnamefont {Schindler}}, \bibinfo
  {author} {\bibfnamefont {D.~S.}\ \bibnamefont {Sanchez}}, \bibinfo {author}
  {\bibfnamefont {I.}~\bibnamefont {Belopolski}}, \bibinfo {author}
  {\bibfnamefont {S.-M.}\ \bibnamefont {Huang}}, \bibinfo {author}
  {\bibfnamefont {B.}~\bibnamefont {Singh}}, \bibinfo {author} {\bibfnamefont
  {D.}~\bibnamefont {Wu}}, \bibinfo {author} {\bibfnamefont {T.-R.}\
  \bibnamefont {Chang}}, \bibinfo {author} {\bibfnamefont {T.}~\bibnamefont
  {Neupert}}, \bibinfo {author} {\bibfnamefont {S.-Y.}\ \bibnamefont {Xu}},
  \bibinfo {author} {\bibfnamefont {H.}~\bibnamefont {Lin}},\ and\ \bibinfo
  {author} {\bibfnamefont {M.~Z.}\ \bibnamefont {Hasan}},\ }\bibfield  {title}
  {\bibinfo {title} {{Topological quantum properties of chiral crystals}},\
  }\href {https://doi.org/10.1038/s41563-018-0169-3} {\bibfield  {journal}
  {\bibinfo  {journal} {Nature Mater.}\ }\textbf {\bibinfo {volume} {17}},\
  \bibinfo {pages} {978} (\bibinfo {year} {2018})}\BibitemShut {NoStop}%
\bibitem [{\citenamefont {Roessli}\ \emph {et~al.}(2002)\citenamefont
  {Roessli}, \citenamefont {B\"oni}, \citenamefont {Fischer},\ and\
  \citenamefont {Endoh}}]{Roessli2002}%
  \BibitemOpen
  \bibfield  {author} {\bibinfo {author} {\bibfnamefont {B.}~\bibnamefont
  {Roessli}}, \bibinfo {author} {\bibfnamefont {P.}~\bibnamefont {B\"oni}},
  \bibinfo {author} {\bibfnamefont {W.~E.}\ \bibnamefont {Fischer}},\ and\
  \bibinfo {author} {\bibfnamefont {Y.}~\bibnamefont {Endoh}},\ }\bibfield
  {title} {\bibinfo {title} {{Chiral Fluctuations in MnSi above the Curie
  Temperature}},\ }\href {https://doi.org/10.1103/PhysRevLett.88.237204}
  {\bibfield  {journal} {\bibinfo  {journal} {Phys. Rev. Lett.}\ }\textbf
  {\bibinfo {volume} {88}},\ \bibinfo {pages} {237204} (\bibinfo {year}
  {2002})}\BibitemShut {NoStop}%
\bibitem [{\citenamefont {Lebrun}\ \emph {et~al.}(2018)\citenamefont {Lebrun},
  \citenamefont {Ross}, \citenamefont {Bender}, \citenamefont {Qaiumzadeh},
  \citenamefont {Baldrati}, \citenamefont {Cramer}, \citenamefont {Brataas},
  \citenamefont {Duine},\ and\ \citenamefont {Kl{\"a}ui}}]{Lebrun2018}%
  \BibitemOpen
  \bibfield  {author} {\bibinfo {author} {\bibfnamefont {R.}~\bibnamefont
  {Lebrun}}, \bibinfo {author} {\bibfnamefont {A.}~\bibnamefont {Ross}},
  \bibinfo {author} {\bibfnamefont {S.~A.}\ \bibnamefont {Bender}}, \bibinfo
  {author} {\bibfnamefont {A.}~\bibnamefont {Qaiumzadeh}}, \bibinfo {author}
  {\bibfnamefont {L.}~\bibnamefont {Baldrati}}, \bibinfo {author}
  {\bibfnamefont {J.}~\bibnamefont {Cramer}}, \bibinfo {author} {\bibfnamefont
  {A.}~\bibnamefont {Brataas}}, \bibinfo {author} {\bibfnamefont {R.~A.}\
  \bibnamefont {Duine}},\ and\ \bibinfo {author} {\bibfnamefont
  {M.}~\bibnamefont {Kl{\"a}ui}},\ }\bibfield  {title} {\bibinfo {title}
  {Tunable long-distance spin transport in a crystalline antiferromagnetic iron
  oxide},\ }\href {https://doi.org/10.1038/s41586-018-0490-7} {\bibfield
  {journal} {\bibinfo  {journal} {Nature}\ }\textbf {\bibinfo {volume} {561}},\
  \bibinfo {pages} {222} (\bibinfo {year} {2018})}\BibitemShut {NoStop}%
\bibitem [{\citenamefont {Nambu}\ \emph {et~al.}(2020)\citenamefont {Nambu},
  \citenamefont {Barker}, \citenamefont {Okino}, \citenamefont {Kikkawa},
  \citenamefont {Shiomi}, \citenamefont {Enderle}, \citenamefont {Weber},
  \citenamefont {Winn}, \citenamefont {Graves-Brook}, \citenamefont
  {Tranquada}, \citenamefont {Ziman}, \citenamefont {Fujita}, \citenamefont
  {Bauer}, \citenamefont {Saitoh},\ and\ \citenamefont {Kakurai}}]{Nambu2020}%
  \BibitemOpen
  \bibfield  {author} {\bibinfo {author} {\bibfnamefont {Y.}~\bibnamefont
  {Nambu}}, \bibinfo {author} {\bibfnamefont {J.}~\bibnamefont {Barker}},
  \bibinfo {author} {\bibfnamefont {Y.}~\bibnamefont {Okino}}, \bibinfo
  {author} {\bibfnamefont {T.}~\bibnamefont {Kikkawa}}, \bibinfo {author}
  {\bibfnamefont {Y.}~\bibnamefont {Shiomi}}, \bibinfo {author} {\bibfnamefont
  {M.}~\bibnamefont {Enderle}}, \bibinfo {author} {\bibfnamefont
  {T.}~\bibnamefont {Weber}}, \bibinfo {author} {\bibfnamefont
  {B.}~\bibnamefont {Winn}}, \bibinfo {author} {\bibfnamefont {M.}~\bibnamefont
  {Graves-Brook}}, \bibinfo {author} {\bibfnamefont {J.~M.}\ \bibnamefont
  {Tranquada}}, \bibinfo {author} {\bibfnamefont {T.}~\bibnamefont {Ziman}},
  \bibinfo {author} {\bibfnamefont {M.}~\bibnamefont {Fujita}}, \bibinfo
  {author} {\bibfnamefont {G.~E.~W.}\ \bibnamefont {Bauer}}, \bibinfo {author}
  {\bibfnamefont {E.}~\bibnamefont {Saitoh}},\ and\ \bibinfo {author}
  {\bibfnamefont {K.}~\bibnamefont {Kakurai}},\ }\bibfield  {title} {\bibinfo
  {title} {{Observation of Magnon Polarization}},\ }\href
  {https://doi.org/10.1103/PhysRevLett.125.027201} {\bibfield  {journal}
  {\bibinfo  {journal} {Phys. Rev. Lett.}\ }\textbf {\bibinfo {volume} {125}},\
  \bibinfo {pages} {027201} (\bibinfo {year} {2020})}\BibitemShut {NoStop}%
\bibitem [{\citenamefont {Liu}\ \emph {et~al.}(2022)\citenamefont {Liu},
  \citenamefont {Xu}, \citenamefont {Liu}, \citenamefont {Zhang}, \citenamefont
  {Meng}, \citenamefont {Sun}, \citenamefont {Gao}, \citenamefont {Zhao},
  \citenamefont {Niu},\ and\ \citenamefont {Li}}]{Liu2022}%
  \BibitemOpen
  \bibfield  {author} {\bibinfo {author} {\bibfnamefont {Y.}~\bibnamefont
  {Liu}}, \bibinfo {author} {\bibfnamefont {Z.}~\bibnamefont {Xu}}, \bibinfo
  {author} {\bibfnamefont {L.}~\bibnamefont {Liu}}, \bibinfo {author}
  {\bibfnamefont {K.}~\bibnamefont {Zhang}}, \bibinfo {author} {\bibfnamefont
  {Y.}~\bibnamefont {Meng}}, \bibinfo {author} {\bibfnamefont {Y.}~\bibnamefont
  {Sun}}, \bibinfo {author} {\bibfnamefont {P.}~\bibnamefont {Gao}}, \bibinfo
  {author} {\bibfnamefont {H.-W.}\ \bibnamefont {Zhao}}, \bibinfo {author}
  {\bibfnamefont {Q.}~\bibnamefont {Niu}},\ and\ \bibinfo {author}
  {\bibfnamefont {J.}~\bibnamefont {Li}},\ }\bibfield  {title} {\bibinfo
  {title} {{Switching magnon chirality in artificial ferrimagnet}},\ }\href
  {https://doi.org/10.1038/s41467-022-28965-7} {\bibfield  {journal} {\bibinfo
  {journal} {Nature Commun.}\ }\textbf {\bibinfo {volume} {13}},\ \bibinfo
  {pages} {1264} (\bibinfo {year} {2022})}\BibitemShut {NoStop}%
\bibitem [{\citenamefont {\v{S}mejkal}\ \emph {et~al.}(2023)\citenamefont
  {\v{S}mejkal}, \citenamefont {Marmodoro}, \citenamefont {Ahn}, \citenamefont
  {Gonz\'alez-Hern\'andez}, \citenamefont {Turek}, \citenamefont {Mankovsky},
  \citenamefont {Ebert}, \citenamefont {D'Souza}, \citenamefont {\v{S}ipr},
  \citenamefont {Sinova},\ and\ \citenamefont {Jungwirth}}]{Smejkal2023}%
  \BibitemOpen
  \bibfield  {author} {\bibinfo {author} {\bibfnamefont {L.}~\bibnamefont
  {\v{S}mejkal}}, \bibinfo {author} {\bibfnamefont {A.}~\bibnamefont
  {Marmodoro}}, \bibinfo {author} {\bibfnamefont {K.-H.}\ \bibnamefont {Ahn}},
  \bibinfo {author} {\bibfnamefont {R.}~\bibnamefont {Gonz\'alez-Hern\'andez}},
  \bibinfo {author} {\bibfnamefont {I.}~\bibnamefont {Turek}}, \bibinfo
  {author} {\bibfnamefont {S.}~\bibnamefont {Mankovsky}}, \bibinfo {author}
  {\bibfnamefont {H.}~\bibnamefont {Ebert}}, \bibinfo {author} {\bibfnamefont
  {S.~W.}\ \bibnamefont {D'Souza}}, \bibinfo {author} {\bibfnamefont
  {O.}~\bibnamefont {\v{S}ipr}}, \bibinfo {author} {\bibfnamefont
  {J.}~\bibnamefont {Sinova}},\ and\ \bibinfo {author} {\bibfnamefont
  {T.}~\bibnamefont {Jungwirth}},\ }\bibfield  {title} {\bibinfo {title}
  {{Chiral Magnons in Altermagnetic ${\mathrm{RuO}}_{2}$}},\ }\href
  {https://doi.org/10.1103/PhysRevLett.131.256703} {\bibfield  {journal}
  {\bibinfo  {journal} {Phys. Rev. Lett.}\ }\textbf {\bibinfo {volume} {131}},\
  \bibinfo {pages} {256703} (\bibinfo {year} {2023})}\BibitemShut {NoStop}%
\bibitem [{\citenamefont {Zhang}\ and\ \citenamefont {Niu}(2015)}]{Zhang2015}%
  \BibitemOpen
  \bibfield  {author} {\bibinfo {author} {\bibfnamefont {L.}~\bibnamefont
  {Zhang}}\ and\ \bibinfo {author} {\bibfnamefont {Q.}~\bibnamefont {Niu}},\
  }\bibfield  {title} {\bibinfo {title} {{Chiral Phonons at High-Symmetry
  Points in Monolayer Hexagonal Lattices}},\ }\href
  {https://doi.org/10.1103/PhysRevLett.115.115502} {\bibfield  {journal}
  {\bibinfo  {journal} {Phys. Rev. Lett.}\ }\textbf {\bibinfo {volume} {115}},\
  \bibinfo {pages} {115502} (\bibinfo {year} {2015})}\BibitemShut {NoStop}%
\bibitem [{\citenamefont {Zhu}\ \emph {et~al.}(2018)\citenamefont {Zhu},
  \citenamefont {Yi}, \citenamefont {Li}, \citenamefont {Xiao}, \citenamefont
  {Zhang}, \citenamefont {Yang}, \citenamefont {Kaindl}, \citenamefont {Li},
  \citenamefont {Wang},\ and\ \citenamefont {Zhang}}]{Zhu2018}%
  \BibitemOpen
  \bibfield  {author} {\bibinfo {author} {\bibfnamefont {H.}~\bibnamefont
  {Zhu}}, \bibinfo {author} {\bibfnamefont {J.}~\bibnamefont {Yi}}, \bibinfo
  {author} {\bibfnamefont {M.-Y.}\ \bibnamefont {Li}}, \bibinfo {author}
  {\bibfnamefont {J.}~\bibnamefont {Xiao}}, \bibinfo {author} {\bibfnamefont
  {L.}~\bibnamefont {Zhang}}, \bibinfo {author} {\bibfnamefont {C.-W.}\
  \bibnamefont {Yang}}, \bibinfo {author} {\bibfnamefont {R.~A.}\ \bibnamefont
  {Kaindl}}, \bibinfo {author} {\bibfnamefont {L.-J.}\ \bibnamefont {Li}},
  \bibinfo {author} {\bibfnamefont {Y.}~\bibnamefont {Wang}},\ and\ \bibinfo
  {author} {\bibfnamefont {X.}~\bibnamefont {Zhang}},\ }\bibfield  {title}
  {\bibinfo {title} {{Observation of chiral phonons}},\ }\href
  {https://doi.org/10.1126/science.aar2711} {\bibfield  {journal} {\bibinfo
  {journal} {Science}\ }\textbf {\bibinfo {volume} {359}},\ \bibinfo {pages}
  {579} (\bibinfo {year} {2018})}\BibitemShut {NoStop}%
\bibitem [{\citenamefont {Juraschek}\ and\ \citenamefont
  {Spaldin}(2019)}]{Juraschek2019}%
  \BibitemOpen
  \bibfield  {author} {\bibinfo {author} {\bibfnamefont {D.~M.}\ \bibnamefont
  {Juraschek}}\ and\ \bibinfo {author} {\bibfnamefont {N.~A.}\ \bibnamefont
  {Spaldin}},\ }\bibfield  {title} {\bibinfo {title} {{Orbital magnetic moments
  of phonons}},\ }\href {https://doi.org/10.1103/PhysRevMaterials.3.064405}
  {\bibfield  {journal} {\bibinfo  {journal} {Phys. Rev. Mater.}\ }\textbf
  {\bibinfo {volume} {3}},\ \bibinfo {pages} {064405} (\bibinfo {year}
  {2019})}\BibitemShut {NoStop}%
\bibitem [{\citenamefont {Wang}\ \emph {et~al.}(2024)\citenamefont {Wang},
  \citenamefont {Sun}, \citenamefont {Li},\ and\ \citenamefont
  {Zhang}}]{Wang2024}%
  \BibitemOpen
  \bibfield  {author} {\bibinfo {author} {\bibfnamefont {T.}~\bibnamefont
  {Wang}}, \bibinfo {author} {\bibfnamefont {H.}~\bibnamefont {Sun}}, \bibinfo
  {author} {\bibfnamefont {X.}~\bibnamefont {Li}},\ and\ \bibinfo {author}
  {\bibfnamefont {L.}~\bibnamefont {Zhang}},\ }\bibfield  {title} {\bibinfo
  {title} {{Chiral Phonons: Prediction, Verification, and Application}},\
  }\href {https://doi.org/10.1021/acs.nanolett.4c00606} {\bibfield  {journal}
  {\bibinfo  {journal} {Nano Lett.}\ }\textbf {\bibinfo {volume} {24}},\
  \bibinfo {pages} {4311} (\bibinfo {year} {2024})}\BibitemShut {NoStop}%
\bibitem [{\citenamefont {Chen}\ \emph {et~al.}(2018)\citenamefont {Chen},
  \citenamefont {Zhang}, \citenamefont {Niu},\ and\ \citenamefont
  {Zhang}}]{Chen2019}%
  \BibitemOpen
  \bibfield  {author} {\bibinfo {author} {\bibfnamefont {H.}~\bibnamefont
  {Chen}}, \bibinfo {author} {\bibfnamefont {W.}~\bibnamefont {Zhang}},
  \bibinfo {author} {\bibfnamefont {Q.}~\bibnamefont {Niu}},\ and\ \bibinfo
  {author} {\bibfnamefont {L.}~\bibnamefont {Zhang}},\ }\bibfield  {title}
  {\bibinfo {title} {{Chiral phonons in two-dimensional materials}},\ }\href
  {https://doi.org/10.1088/2053-1583/aaf292} {\bibfield  {journal} {\bibinfo
  {journal} {2D Materials}\ }\textbf {\bibinfo {volume} {6}},\ \bibinfo {pages}
  {012002} (\bibinfo {year} {2018})}\BibitemShut {NoStop}%
\bibitem [{\citenamefont {Zhang}\ and\ \citenamefont {Niu}(2014)}]{Zhang2014}%
  \BibitemOpen
  \bibfield  {author} {\bibinfo {author} {\bibfnamefont {L.}~\bibnamefont
  {Zhang}}\ and\ \bibinfo {author} {\bibfnamefont {Q.}~\bibnamefont {Niu}},\
  }\bibfield  {title} {\bibinfo {title} {{Angular Momentum of Phonons and the
  {E}instein--de {H}aas Effect}},\ }\href
  {https://doi.org/10.1103/PhysRevLett.112.085503} {\bibfield  {journal}
  {\bibinfo  {journal} {Phys. Rev. Lett.}\ }\textbf {\bibinfo {volume} {112}},\
  \bibinfo {pages} {085503} (\bibinfo {year} {2014})}\BibitemShut {NoStop}%
\bibitem [{\citenamefont {Strohm}\ \emph {et~al.}(2005)\citenamefont {Strohm},
  \citenamefont {Rikken},\ and\ \citenamefont {Wyder}}]{Strohm2005}%
  \BibitemOpen
  \bibfield  {author} {\bibinfo {author} {\bibfnamefont {C.}~\bibnamefont
  {Strohm}}, \bibinfo {author} {\bibfnamefont {G.~L. J.~A.}\ \bibnamefont
  {Rikken}},\ and\ \bibinfo {author} {\bibfnamefont {P.}~\bibnamefont
  {Wyder}},\ }\bibfield  {title} {\bibinfo {title} {{Phenomenological Evidence
  for the Phonon {H}all Effect}},\ }\href
  {https://doi.org/10.1103/PhysRevLett.95.155901} {\bibfield  {journal}
  {\bibinfo  {journal} {Phys. Rev. Lett.}\ }\textbf {\bibinfo {volume} {95}},\
  \bibinfo {pages} {155901} (\bibinfo {year} {2005})}\BibitemShut {NoStop}%
\bibitem [{\citenamefont {Grissonnanche}\ \emph {et~al.}(2020)\citenamefont
  {Grissonnanche}, \citenamefont {Th{\'e}riault}, \citenamefont {Gourgout},
  \citenamefont {Boulanger}, \citenamefont {Lefran{\c{c}}ois}, \citenamefont
  {Ataei}, \citenamefont {Lalibert{\'e}}, \citenamefont {Dion}, \citenamefont
  {Zhou}, \citenamefont {Pyon}, \citenamefont {Takayama}, \citenamefont
  {Takagi}, \citenamefont {Doiron-Leyraud},\ and\ \citenamefont
  {Taillefer}}]{Grissonnanche2020}%
  \BibitemOpen
  \bibfield  {author} {\bibinfo {author} {\bibfnamefont {G.}~\bibnamefont
  {Grissonnanche}}, \bibinfo {author} {\bibfnamefont {S.}~\bibnamefont
  {Th{\'e}riault}}, \bibinfo {author} {\bibfnamefont {A.}~\bibnamefont
  {Gourgout}}, \bibinfo {author} {\bibfnamefont {M.-E.}\ \bibnamefont
  {Boulanger}}, \bibinfo {author} {\bibfnamefont {E.}~\bibnamefont
  {Lefran{\c{c}}ois}}, \bibinfo {author} {\bibfnamefont {A.}~\bibnamefont
  {Ataei}}, \bibinfo {author} {\bibfnamefont {F.}~\bibnamefont
  {Lalibert{\'e}}}, \bibinfo {author} {\bibfnamefont {M.}~\bibnamefont {Dion}},
  \bibinfo {author} {\bibfnamefont {J.-S.}\ \bibnamefont {Zhou}}, \bibinfo
  {author} {\bibfnamefont {S.}~\bibnamefont {Pyon}}, \bibinfo {author}
  {\bibfnamefont {T.}~\bibnamefont {Takayama}}, \bibinfo {author}
  {\bibfnamefont {H.}~\bibnamefont {Takagi}}, \bibinfo {author} {\bibfnamefont
  {N.}~\bibnamefont {Doiron-Leyraud}},\ and\ \bibinfo {author} {\bibfnamefont
  {L.}~\bibnamefont {Taillefer}},\ }\bibfield  {title} {\bibinfo {title}
  {{Chiral phonons in the pseudogap phase of cuprates}},\ }\href
  {https://doi.org/10.1038/s41567-020-0965-y} {\bibfield  {journal} {\bibinfo
  {journal} {Nature Phys.}\ }\textbf {\bibinfo {volume} {16}},\ \bibinfo
  {pages} {1108} (\bibinfo {year} {2020})}\BibitemShut {NoStop}%
\bibitem [{\citenamefont {Park}\ and\ \citenamefont {Yang}(2020)}]{Park2020b}%
  \BibitemOpen
  \bibfield  {author} {\bibinfo {author} {\bibfnamefont {S.}~\bibnamefont
  {Park}}\ and\ \bibinfo {author} {\bibfnamefont {B.-J.}\ \bibnamefont
  {Yang}},\ }\bibfield  {title} {\bibinfo {title} {{Phonon Angular Momentum
  Hall Effect}},\ }\href {https://doi.org/10.1021/acs.nanolett.0c03220}
  {\bibfield  {journal} {\bibinfo  {journal} {Nano Lett.}\ }\textbf {\bibinfo
  {volume} {20}},\ \bibinfo {pages} {7694} (\bibinfo {year}
  {2020})}\BibitemShut {NoStop}%
\bibitem [{\citenamefont {Flebus}\ and\ \citenamefont
  {MacDonald}(2023)}]{Flebus2023}%
  \BibitemOpen
  \bibfield  {author} {\bibinfo {author} {\bibfnamefont {B.}~\bibnamefont
  {Flebus}}\ and\ \bibinfo {author} {\bibfnamefont {A.~H.}\ \bibnamefont
  {MacDonald}},\ }\bibfield  {title} {\bibinfo {title} {{Phonon Hall Viscosity
  of Ionic Crystals}},\ }\href {https://doi.org/10.1103/PhysRevLett.131.236301}
  {\bibfield  {journal} {\bibinfo  {journal} {Phys. Rev. Lett.}\ }\textbf
  {\bibinfo {volume} {131}},\ \bibinfo {pages} {236301} (\bibinfo {year}
  {2023})}\BibitemShut {NoStop}%
\bibitem [{\citenamefont {Dornes}\ \emph {et~al.}(2019)\citenamefont {Dornes},
  \citenamefont {Acremann}, \citenamefont {Savoini}, \citenamefont {Kubli},
  \citenamefont {Neugebauer}, \citenamefont {Abreu}, \citenamefont {Huber},
  \citenamefont {Lantz}, \citenamefont {Vaz}, \citenamefont {Lemke},
  \citenamefont {Bothschafter}, \citenamefont {Porer}, \citenamefont
  {Esposito}, \citenamefont {Rettig}, \citenamefont {Buzzi}, \citenamefont
  {Alberca}, \citenamefont {Windsor}, \citenamefont {Beaud}, \citenamefont
  {Staub}, \citenamefont {Zhu}, \citenamefont {Song}, \citenamefont {Glownia},\
  and\ \citenamefont {Johnson}}]{Dornes2019}%
  \BibitemOpen
  \bibfield  {author} {\bibinfo {author} {\bibfnamefont {C.}~\bibnamefont
  {Dornes}}, \bibinfo {author} {\bibfnamefont {Y.}~\bibnamefont {Acremann}},
  \bibinfo {author} {\bibfnamefont {M.}~\bibnamefont {Savoini}}, \bibinfo
  {author} {\bibfnamefont {M.}~\bibnamefont {Kubli}}, \bibinfo {author}
  {\bibfnamefont {M.~J.}\ \bibnamefont {Neugebauer}}, \bibinfo {author}
  {\bibfnamefont {E.}~\bibnamefont {Abreu}}, \bibinfo {author} {\bibfnamefont
  {L.}~\bibnamefont {Huber}}, \bibinfo {author} {\bibfnamefont
  {G.}~\bibnamefont {Lantz}}, \bibinfo {author} {\bibfnamefont {C.~A.~F.}\
  \bibnamefont {Vaz}}, \bibinfo {author} {\bibfnamefont {H.}~\bibnamefont
  {Lemke}}, \bibinfo {author} {\bibfnamefont {E.~M.}\ \bibnamefont
  {Bothschafter}}, \bibinfo {author} {\bibfnamefont {M.}~\bibnamefont {Porer}},
  \bibinfo {author} {\bibfnamefont {V.}~\bibnamefont {Esposito}}, \bibinfo
  {author} {\bibfnamefont {L.}~\bibnamefont {Rettig}}, \bibinfo {author}
  {\bibfnamefont {M.}~\bibnamefont {Buzzi}}, \bibinfo {author} {\bibfnamefont
  {A.}~\bibnamefont {Alberca}}, \bibinfo {author} {\bibfnamefont {Y.~W.}\
  \bibnamefont {Windsor}}, \bibinfo {author} {\bibfnamefont {P.}~\bibnamefont
  {Beaud}}, \bibinfo {author} {\bibfnamefont {U.}~\bibnamefont {Staub}},
  \bibinfo {author} {\bibfnamefont {D.}~\bibnamefont {Zhu}}, \bibinfo {author}
  {\bibfnamefont {S.}~\bibnamefont {Song}}, \bibinfo {author} {\bibfnamefont
  {J.~M.}\ \bibnamefont {Glownia}},\ and\ \bibinfo {author} {\bibfnamefont
  {S.~L.}\ \bibnamefont {Johnson}},\ }\bibfield  {title} {\bibinfo {title}
  {{The ultrafast Einstein--de Haas effect}},\ }\href
  {https://doi.org/10.1038/s41586-018-0822-7} {\bibfield  {journal} {\bibinfo
  {journal} {Nature}\ }\textbf {\bibinfo {volume} {565}},\ \bibinfo {pages}
  {209} (\bibinfo {year} {2019})}\BibitemShut {NoStop}%
\bibitem [{\citenamefont {Tauchert}\ \emph {et~al.}(2022)\citenamefont
  {Tauchert}, \citenamefont {Volkov}, \citenamefont {Ehberger}, \citenamefont
  {Kazenwadel}, \citenamefont {Evers}, \citenamefont {Lange}, \citenamefont
  {Donges}, \citenamefont {Book}, \citenamefont {Kreuzpaintner}, \citenamefont
  {Nowak},\ and\ \citenamefont {Baum}}]{Tauchert2022}%
  \BibitemOpen
  \bibfield  {author} {\bibinfo {author} {\bibfnamefont {S.~R.}\ \bibnamefont
  {Tauchert}}, \bibinfo {author} {\bibfnamefont {M.}~\bibnamefont {Volkov}},
  \bibinfo {author} {\bibfnamefont {D.}~\bibnamefont {Ehberger}}, \bibinfo
  {author} {\bibfnamefont {D.}~\bibnamefont {Kazenwadel}}, \bibinfo {author}
  {\bibfnamefont {M.}~\bibnamefont {Evers}}, \bibinfo {author} {\bibfnamefont
  {H.}~\bibnamefont {Lange}}, \bibinfo {author} {\bibfnamefont
  {A.}~\bibnamefont {Donges}}, \bibinfo {author} {\bibfnamefont
  {A.}~\bibnamefont {Book}}, \bibinfo {author} {\bibfnamefont {W.}~\bibnamefont
  {Kreuzpaintner}}, \bibinfo {author} {\bibfnamefont {U.}~\bibnamefont
  {Nowak}},\ and\ \bibinfo {author} {\bibfnamefont {P.}~\bibnamefont {Baum}},\
  }\bibfield  {title} {\bibinfo {title} {{Polarized phonons carry angular
  momentum in ultrafast demagnetization}},\ }\href
  {https://doi.org/10.1038/s41586-021-04306-4} {\bibfield  {journal} {\bibinfo
  {journal} {Nature}\ }\textbf {\bibinfo {volume} {602}},\ \bibinfo {pages}
  {73} (\bibinfo {year} {2022})}\BibitemShut {NoStop}%
\bibitem [{\citenamefont {Luo}\ \emph {et~al.}(2023)\citenamefont {Luo},
  \citenamefont {Lin}, \citenamefont {Zhang}, \citenamefont {Chen},
  \citenamefont {Blackert}, \citenamefont {Xu}, \citenamefont {Yakobson},\ and\
  \citenamefont {Zhu}}]{Luo2023}%
  \BibitemOpen
  \bibfield  {author} {\bibinfo {author} {\bibfnamefont {J.}~\bibnamefont
  {Luo}}, \bibinfo {author} {\bibfnamefont {T.}~\bibnamefont {Lin}}, \bibinfo
  {author} {\bibfnamefont {J.}~\bibnamefont {Zhang}}, \bibinfo {author}
  {\bibfnamefont {X.}~\bibnamefont {Chen}}, \bibinfo {author} {\bibfnamefont
  {E.~R.}\ \bibnamefont {Blackert}}, \bibinfo {author} {\bibfnamefont
  {R.}~\bibnamefont {Xu}}, \bibinfo {author} {\bibfnamefont {B.~I.}\
  \bibnamefont {Yakobson}},\ and\ \bibinfo {author} {\bibfnamefont
  {H.}~\bibnamefont {Zhu}},\ }\bibfield  {title} {\bibinfo {title} {{Large
  effective magnetic fields from chiral phonons in rare-earth halides}},\
  }\href {https://doi.org/10.1126/science.adi9601} {\bibfield  {journal}
  {\bibinfo  {journal} {Science}\ }\textbf {\bibinfo {volume} {382}},\ \bibinfo
  {pages} {698} (\bibinfo {year} {2023})}\BibitemShut {NoStop}%
\bibitem [{\citenamefont {Davies}\ \emph {et~al.}(2024)\citenamefont {Davies},
  \citenamefont {Fennema}, \citenamefont {Tsukamoto}, \citenamefont
  {Razdolski}, \citenamefont {Kimel},\ and\ \citenamefont
  {Kirilyuk}}]{Davies2024}%
  \BibitemOpen
  \bibfield  {author} {\bibinfo {author} {\bibfnamefont {C.~S.}\ \bibnamefont
  {Davies}}, \bibinfo {author} {\bibfnamefont {F.~G.~N.}\ \bibnamefont
  {Fennema}}, \bibinfo {author} {\bibfnamefont {A.}~\bibnamefont {Tsukamoto}},
  \bibinfo {author} {\bibfnamefont {I.}~\bibnamefont {Razdolski}}, \bibinfo
  {author} {\bibfnamefont {A.~V.}\ \bibnamefont {Kimel}},\ and\ \bibinfo
  {author} {\bibfnamefont {A.}~\bibnamefont {Kirilyuk}},\ }\bibfield  {title}
  {\bibinfo {title} {{Phononic switching of magnetization by the ultrafast
  Barnett effect}},\ }\href {https://doi.org/10.1038/s41586-024-07200-x}
  {\bibfield  {journal} {\bibinfo  {journal} {Nature}\ }\textbf {\bibinfo
  {volume} {628}},\ \bibinfo {pages} {540} (\bibinfo {year}
  {2024})}\BibitemShut {NoStop}%
\bibitem [{\citenamefont {Holanda}\ \emph {et~al.}(2018)\citenamefont
  {Holanda}, \citenamefont {Maior}, \citenamefont {Azevedo},\ and\
  \citenamefont {Rezende}}]{Holanda2018}%
  \BibitemOpen
  \bibfield  {author} {\bibinfo {author} {\bibfnamefont {J.}~\bibnamefont
  {Holanda}}, \bibinfo {author} {\bibfnamefont {D.~S.}\ \bibnamefont {Maior}},
  \bibinfo {author} {\bibfnamefont {A.}~\bibnamefont {Azevedo}},\ and\ \bibinfo
  {author} {\bibfnamefont {S.~M.}\ \bibnamefont {Rezende}},\ }\bibfield
  {title} {\bibinfo {title} {{Detecting the phonon spin in magnon--phonon
  conversion experiments}},\ }\href {https://doi.org/10.1038/s41567-018-0079-y}
  {\bibfield  {journal} {\bibinfo  {journal} {Nature Phys.}\ }\textbf {\bibinfo
  {volume} {14}},\ \bibinfo {pages} {500} (\bibinfo {year} {2018})}\BibitemShut
  {NoStop}%
\bibitem [{\citenamefont {Juraschek}\ \emph {et~al.}(2017)\citenamefont
  {Juraschek}, \citenamefont {Fechner}, \citenamefont {Balatsky},\ and\
  \citenamefont {Spaldin}}]{Juraschek2017}%
  \BibitemOpen
  \bibfield  {author} {\bibinfo {author} {\bibfnamefont {D.~M.}\ \bibnamefont
  {Juraschek}}, \bibinfo {author} {\bibfnamefont {M.}~\bibnamefont {Fechner}},
  \bibinfo {author} {\bibfnamefont {A.~V.}\ \bibnamefont {Balatsky}},\ and\
  \bibinfo {author} {\bibfnamefont {N.~A.}\ \bibnamefont {Spaldin}},\
  }\bibfield  {title} {\bibinfo {title} {{Dynamical multiferroicity}},\ }\href
  {https://doi.org/10.1103/PhysRevMaterials.1.014401} {\bibfield  {journal}
  {\bibinfo  {journal} {Phys. Rev. Mater.}\ }\textbf {\bibinfo {volume} {1}},\
  \bibinfo {pages} {014401} (\bibinfo {year} {2017})}\BibitemShut {NoStop}%
\bibitem [{\citenamefont {Ren}\ \emph {et~al.}(2021)\citenamefont {Ren},
  \citenamefont {Xiao}, \citenamefont {Saparov},\ and\ \citenamefont
  {Niu}}]{Ren2021}%
  \BibitemOpen
  \bibfield  {author} {\bibinfo {author} {\bibfnamefont {Y.}~\bibnamefont
  {Ren}}, \bibinfo {author} {\bibfnamefont {C.}~\bibnamefont {Xiao}}, \bibinfo
  {author} {\bibfnamefont {D.}~\bibnamefont {Saparov}},\ and\ \bibinfo {author}
  {\bibfnamefont {Q.}~\bibnamefont {Niu}},\ }\bibfield  {title} {\bibinfo
  {title} {{Phonon Magnetic Moment from Electronic Topological
  Magnetization}},\ }\href {https://doi.org/10.1103/PhysRevLett.127.186403}
  {\bibfield  {journal} {\bibinfo  {journal} {Phys. Rev. Lett.}\ }\textbf
  {\bibinfo {volume} {127}},\ \bibinfo {pages} {186403} (\bibinfo {year}
  {2021})}\BibitemShut {NoStop}%
\bibitem [{\citenamefont {Basini}\ \emph {et~al.}(2024)\citenamefont {Basini},
  \citenamefont {Pancaldi}, \citenamefont {Wehinger}, \citenamefont {Udina},
  \citenamefont {Unikandanunni}, \citenamefont {Tadano}, \citenamefont
  {Hoffmann}, \citenamefont {Balatsky},\ and\ \citenamefont
  {Bonetti}}]{Basini2024}%
  \BibitemOpen
  \bibfield  {author} {\bibinfo {author} {\bibfnamefont {M.}~\bibnamefont
  {Basini}}, \bibinfo {author} {\bibfnamefont {M.}~\bibnamefont {Pancaldi}},
  \bibinfo {author} {\bibfnamefont {B.}~\bibnamefont {Wehinger}}, \bibinfo
  {author} {\bibfnamefont {M.}~\bibnamefont {Udina}}, \bibinfo {author}
  {\bibfnamefont {V.}~\bibnamefont {Unikandanunni}}, \bibinfo {author}
  {\bibfnamefont {T.}~\bibnamefont {Tadano}}, \bibinfo {author} {\bibfnamefont
  {M.~C.}\ \bibnamefont {Hoffmann}}, \bibinfo {author} {\bibfnamefont {A.~V.}\
  \bibnamefont {Balatsky}},\ and\ \bibinfo {author} {\bibfnamefont
  {S.}~\bibnamefont {Bonetti}},\ }\bibfield  {title} {\bibinfo {title}
  {{Terahertz electric-field-driven dynamical multiferroicity in SrTiO$_3$}},\
  }\href {https://doi.org/10.1038/s41586-024-07175-9} {\bibfield  {journal}
  {\bibinfo  {journal} {Nature}\ }\textbf {\bibinfo {volume} {628}},\ \bibinfo
  {pages} {534} (\bibinfo {year} {2024})}\BibitemShut {NoStop}%
\bibitem [{\citenamefont {Cheong}\ and\ \citenamefont {Xu}(2022)}]{Cheong2022}%
  \BibitemOpen
  \bibfield  {author} {\bibinfo {author} {\bibfnamefont {S.-W.}\ \bibnamefont
  {Cheong}}\ and\ \bibinfo {author} {\bibfnamefont {X.}~\bibnamefont {Xu}},\
  }\bibfield  {title} {\bibinfo {title} {{Magnetic chirality}},\ }\href
  {https://doi.org/10.1038/s41535-022-00447-5} {\bibfield  {journal} {\bibinfo
  {journal} {npj Quantum Mater.}\ }\textbf {\bibinfo {volume} {7}},\ \bibinfo
  {pages} {40} (\bibinfo {year} {2022})}\BibitemShut {NoStop}%
\bibitem [{\citenamefont {Ueda}\ \emph {et~al.}(2023)\citenamefont {Ueda},
  \citenamefont {Garc{\'i}a-Fern{\'a}ndez}, \citenamefont {Agrestini},
  \citenamefont {Romao}, \citenamefont {van~den Brink}, \citenamefont
  {Spaldin}, \citenamefont {Zhou},\ and\ \citenamefont {Staub}}]{Ueda2023}%
  \BibitemOpen
  \bibfield  {author} {\bibinfo {author} {\bibfnamefont {H.}~\bibnamefont
  {Ueda}}, \bibinfo {author} {\bibfnamefont {M.}~\bibnamefont
  {Garc{\'i}a-Fern{\'a}ndez}}, \bibinfo {author} {\bibfnamefont
  {S.}~\bibnamefont {Agrestini}}, \bibinfo {author} {\bibfnamefont {C.~P.}\
  \bibnamefont {Romao}}, \bibinfo {author} {\bibfnamefont {J.}~\bibnamefont
  {van~den Brink}}, \bibinfo {author} {\bibfnamefont {N.~A.}\ \bibnamefont
  {Spaldin}}, \bibinfo {author} {\bibfnamefont {K.-J.}\ \bibnamefont {Zhou}},\
  and\ \bibinfo {author} {\bibfnamefont {U.}~\bibnamefont {Staub}},\ }\bibfield
   {title} {\bibinfo {title} {{Chiral phonons in quartz probed by X-rays}},\
  }\href {https://doi.org/10.1038/s41586-023-06016-5} {\bibfield  {journal}
  {\bibinfo  {journal} {Nature}\ }\textbf {\bibinfo {volume} {618}},\ \bibinfo
  {pages} {946} (\bibinfo {year} {2023})}\BibitemShut {NoStop}%
\bibitem [{\citenamefont {Barron}(2004)}]{Barron_2004}%
  \BibitemOpen
  \bibfield  {author} {\bibinfo {author} {\bibfnamefont {L.~D.}\ \bibnamefont
  {Barron}},\ }\href@noop {} {\emph {\bibinfo {title} {Molecular Light
  Scattering and Optical Activity}}},\ \bibinfo {edition} {2nd}\ ed.\ (\bibinfo
   {publisher} {Cambridge University Press},\ \bibinfo {year}
  {2004})\BibitemShut {NoStop}%
\bibitem [{\citenamefont {Juraschek}\ \emph {et~al.}(2025)\citenamefont
  {Juraschek}, \citenamefont {Geilhufe}, \citenamefont {Zhu}, \citenamefont
  {Basini}, \citenamefont {Baum}, \citenamefont {Baydin}, \citenamefont
  {Chaudhary}, \citenamefont {Fechner}, \citenamefont {Flebus}, \citenamefont
  {Grissonnanche}, \citenamefont {Kirilyuk}, \citenamefont {Lemeshko},
  \citenamefont {Maehrlein}, \citenamefont {Mignolet}, \citenamefont
  {Murakami}, \citenamefont {Niu}, \citenamefont {Nowak}, \citenamefont
  {Romao}, \citenamefont {Rostami}, \citenamefont {Satoh}, \citenamefont
  {Spaldin}, \citenamefont {Ueda},\ and\ \citenamefont
  {Zhang}}]{Juraschek2025}%
  \BibitemOpen
  \bibfield  {author} {\bibinfo {author} {\bibfnamefont {D.~M.}\ \bibnamefont
  {Juraschek}}, \bibinfo {author} {\bibfnamefont {R.~M.}\ \bibnamefont
  {Geilhufe}}, \bibinfo {author} {\bibfnamefont {H.}~\bibnamefont {Zhu}},
  \bibinfo {author} {\bibfnamefont {M.}~\bibnamefont {Basini}}, \bibinfo
  {author} {\bibfnamefont {P.}~\bibnamefont {Baum}}, \bibinfo {author}
  {\bibfnamefont {A.}~\bibnamefont {Baydin}}, \bibinfo {author} {\bibfnamefont
  {S.}~\bibnamefont {Chaudhary}}, \bibinfo {author} {\bibfnamefont
  {M.}~\bibnamefont {Fechner}}, \bibinfo {author} {\bibfnamefont
  {B.}~\bibnamefont {Flebus}}, \bibinfo {author} {\bibfnamefont
  {G.}~\bibnamefont {Grissonnanche}}, \bibinfo {author} {\bibfnamefont {A.~I.}\
  \bibnamefont {Kirilyuk}}, \bibinfo {author} {\bibfnamefont {M.}~\bibnamefont
  {Lemeshko}}, \bibinfo {author} {\bibfnamefont {S.~F.}\ \bibnamefont
  {Maehrlein}}, \bibinfo {author} {\bibfnamefont {M.}~\bibnamefont {Mignolet}},
  \bibinfo {author} {\bibfnamefont {S.}~\bibnamefont {Murakami}}, \bibinfo
  {author} {\bibfnamefont {Q.}~\bibnamefont {Niu}}, \bibinfo {author}
  {\bibfnamefont {U.}~\bibnamefont {Nowak}}, \bibinfo {author} {\bibfnamefont
  {C.~P.}\ \bibnamefont {Romao}}, \bibinfo {author} {\bibfnamefont
  {H.}~\bibnamefont {Rostami}}, \bibinfo {author} {\bibfnamefont
  {T.}~\bibnamefont {Satoh}}, \bibinfo {author} {\bibfnamefont {N.~A.}\
  \bibnamefont {Spaldin}}, \bibinfo {author} {\bibfnamefont {H.}~\bibnamefont
  {Ueda}},\ and\ \bibinfo {author} {\bibfnamefont {L.}~\bibnamefont {Zhang}},\
  }\bibfield  {title} {\bibinfo {title} {Chiral phonons},\ }\href
  {https://doi.org/10.1038/s41567-025-03001-9} {\bibfield  {journal} {\bibinfo
  {journal} {Nature Physics}\ }\textbf {\bibinfo {volume} {21}},\ \bibinfo
  {pages} {1532} (\bibinfo {year} {2025})}\BibitemShut {NoStop}%
\bibitem [{\citenamefont {Ishito}\ \emph
  {et~al.}(2023{\natexlab{a}})\citenamefont {Ishito}, \citenamefont {Mao},
  \citenamefont {Kousaka}, \citenamefont {Togawa}, \citenamefont {Iwasaki},
  \citenamefont {Zhang}, \citenamefont {Murakami}, \citenamefont {Kishine},\
  and\ \citenamefont {Satoh}}]{Ishito2023}%
  \BibitemOpen
  \bibfield  {author} {\bibinfo {author} {\bibfnamefont {K.}~\bibnamefont
  {Ishito}}, \bibinfo {author} {\bibfnamefont {H.}~\bibnamefont {Mao}},
  \bibinfo {author} {\bibfnamefont {Y.}~\bibnamefont {Kousaka}}, \bibinfo
  {author} {\bibfnamefont {Y.}~\bibnamefont {Togawa}}, \bibinfo {author}
  {\bibfnamefont {S.}~\bibnamefont {Iwasaki}}, \bibinfo {author} {\bibfnamefont
  {T.}~\bibnamefont {Zhang}}, \bibinfo {author} {\bibfnamefont
  {S.}~\bibnamefont {Murakami}}, \bibinfo {author} {\bibfnamefont {J.-i.}\
  \bibnamefont {Kishine}},\ and\ \bibinfo {author} {\bibfnamefont
  {T.}~\bibnamefont {Satoh}},\ }\bibfield  {title} {\bibinfo {title} {{Truly
  chiral phonons in $\alpha$-{HgS}}},\ }\href
  {https://doi.org/10.1038/s41567-022-01790-x} {\bibfield  {journal} {\bibinfo
  {journal} {Nature Phys.}\ }\textbf {\bibinfo {volume} {19}},\ \bibinfo
  {pages} {35} (\bibinfo {year} {2023}{\natexlab{a}})}\BibitemShut {NoStop}%
\bibitem [{\citenamefont {Ishito}\ \emph
  {et~al.}(2023{\natexlab{b}})\citenamefont {Ishito}, \citenamefont {Mao},
  \citenamefont {Kobayashi}, \citenamefont {Kousaka}, \citenamefont {Togawa},
  \citenamefont {Kusunose}, \citenamefont {Kishine},\ and\ \citenamefont
  {Satoh}}]{Ishito2023b}%
  \BibitemOpen
  \bibfield  {author} {\bibinfo {author} {\bibfnamefont {K.}~\bibnamefont
  {Ishito}}, \bibinfo {author} {\bibfnamefont {H.}~\bibnamefont {Mao}},
  \bibinfo {author} {\bibfnamefont {K.}~\bibnamefont {Kobayashi}}, \bibinfo
  {author} {\bibfnamefont {Y.}~\bibnamefont {Kousaka}}, \bibinfo {author}
  {\bibfnamefont {Y.}~\bibnamefont {Togawa}}, \bibinfo {author} {\bibfnamefont
  {H.}~\bibnamefont {Kusunose}}, \bibinfo {author} {\bibfnamefont {J.-i.}\
  \bibnamefont {Kishine}},\ and\ \bibinfo {author} {\bibfnamefont
  {T.}~\bibnamefont {Satoh}},\ }\bibfield  {title} {\bibinfo {title} {Chiral
  phonons: circularly polarized raman spectroscopy and ab initio calculations
  in a chiral crystal tellurium},\ }\href
  {https://doi.org/https://doi.org/10.1002/chir.23544} {\bibfield  {journal}
  {\bibinfo  {journal} {Chirality}\ }\textbf {\bibinfo {volume} {35}},\
  \bibinfo {pages} {338} (\bibinfo {year} {2023}{\natexlab{b}})},\ \Eprint
  {https://arxiv.org/abs/https://onlinelibrary.wiley.com/doi/pdf/10.1002/chir.23544}
  {https://onlinelibrary.wiley.com/doi/pdf/10.1002/chir.23544} \BibitemShut
  {NoStop}%
\bibitem [{\citenamefont {Kim}\ \emph {et~al.}(2023)\citenamefont {Kim},
  \citenamefont {Vetter}, \citenamefont {Yan}, \citenamefont {Yang},
  \citenamefont {Wang}, \citenamefont {Sun}, \citenamefont {Yang},
  \citenamefont {Comstock}, \citenamefont {Li}, \citenamefont {Zhou},
  \citenamefont {Zhang}, \citenamefont {You}, \citenamefont {Sun},\ and\
  \citenamefont {Liu}}]{Kim2023}%
  \BibitemOpen
  \bibfield  {author} {\bibinfo {author} {\bibfnamefont {K.}~\bibnamefont
  {Kim}}, \bibinfo {author} {\bibfnamefont {E.}~\bibnamefont {Vetter}},
  \bibinfo {author} {\bibfnamefont {L.}~\bibnamefont {Yan}}, \bibinfo {author}
  {\bibfnamefont {C.}~\bibnamefont {Yang}}, \bibinfo {author} {\bibfnamefont
  {Z.}~\bibnamefont {Wang}}, \bibinfo {author} {\bibfnamefont {R.}~\bibnamefont
  {Sun}}, \bibinfo {author} {\bibfnamefont {Y.}~\bibnamefont {Yang}}, \bibinfo
  {author} {\bibfnamefont {A.~H.}\ \bibnamefont {Comstock}}, \bibinfo {author}
  {\bibfnamefont {X.}~\bibnamefont {Li}}, \bibinfo {author} {\bibfnamefont
  {J.}~\bibnamefont {Zhou}}, \bibinfo {author} {\bibfnamefont {L.}~\bibnamefont
  {Zhang}}, \bibinfo {author} {\bibfnamefont {W.}~\bibnamefont {You}}, \bibinfo
  {author} {\bibfnamefont {D.}~\bibnamefont {Sun}},\ and\ \bibinfo {author}
  {\bibfnamefont {J.}~\bibnamefont {Liu}},\ }\bibfield  {title} {\bibinfo
  {title} {Chiral-phonon-activated spin seebeck effect},\ }\href
  {https://doi.org/10.1038/s41563-023-01473-9} {\bibfield  {journal} {\bibinfo
  {journal} {Nature Materials}\ }\textbf {\bibinfo {volume} {22}},\ \bibinfo
  {pages} {322} (\bibinfo {year} {2023})}\BibitemShut {NoStop}%
\bibitem [{\citenamefont {Ohe}\ \emph {et~al.}(2024)\citenamefont {Ohe},
  \citenamefont {Shishido}, \citenamefont {Kato}, \citenamefont {Utsumi},
  \citenamefont {Matsuura},\ and\ \citenamefont {Togawa}}]{Ohe2024}%
  \BibitemOpen
  \bibfield  {author} {\bibinfo {author} {\bibfnamefont {K.}~\bibnamefont
  {Ohe}}, \bibinfo {author} {\bibfnamefont {H.}~\bibnamefont {Shishido}},
  \bibinfo {author} {\bibfnamefont {M.}~\bibnamefont {Kato}}, \bibinfo {author}
  {\bibfnamefont {S.}~\bibnamefont {Utsumi}}, \bibinfo {author} {\bibfnamefont
  {H.}~\bibnamefont {Matsuura}},\ and\ \bibinfo {author} {\bibfnamefont
  {Y.}~\bibnamefont {Togawa}},\ }\bibfield  {title} {\bibinfo {title}
  {Chirality-induced selectivity of phonon angular momenta in chiral quartz
  crystals},\ }\href {https://doi.org/10.1103/PhysRevLett.132.056302}
  {\bibfield  {journal} {\bibinfo  {journal} {Phys. Rev. Lett.}\ }\textbf
  {\bibinfo {volume} {132}},\ \bibinfo {pages} {056302} (\bibinfo {year}
  {2024})}\BibitemShut {NoStop}%
\bibitem [{\citenamefont {Zhang}\ \emph {et~al.}(2025)\citenamefont {Zhang},
  \citenamefont {Peshcherenko}, \citenamefont {Yang}, \citenamefont {Ward},
  \citenamefont {Raghuvanshi}, \citenamefont {Lindsay}, \citenamefont {Felser},
  \citenamefont {Zhang}, \citenamefont {Yan},\ and\ \citenamefont
  {Miao}}]{Zhang2025}%
  \BibitemOpen
  \bibfield  {author} {\bibinfo {author} {\bibfnamefont {H.}~\bibnamefont
  {Zhang}}, \bibinfo {author} {\bibfnamefont {N.}~\bibnamefont {Peshcherenko}},
  \bibinfo {author} {\bibfnamefont {F.}~\bibnamefont {Yang}}, \bibinfo {author}
  {\bibfnamefont {T.~Z.}\ \bibnamefont {Ward}}, \bibinfo {author}
  {\bibfnamefont {P.}~\bibnamefont {Raghuvanshi}}, \bibinfo {author}
  {\bibfnamefont {L.}~\bibnamefont {Lindsay}}, \bibinfo {author} {\bibfnamefont
  {C.}~\bibnamefont {Felser}}, \bibinfo {author} {\bibfnamefont
  {Y.}~\bibnamefont {Zhang}}, \bibinfo {author} {\bibfnamefont {J.-Q.}\
  \bibnamefont {Yan}},\ and\ \bibinfo {author} {\bibfnamefont {H.}~\bibnamefont
  {Miao}},\ }\bibfield  {title} {\bibinfo {title} {Measurement of phonon
  angular momentum},\ }\href {https://doi.org/10.1038/s41567-025-02952-3}
  {\bibfield  {journal} {\bibinfo  {journal} {Nature Physics}\ }\textbf
  {\bibinfo {volume} {21}},\ \bibinfo {pages} {1387} (\bibinfo {year}
  {2025})}\BibitemShut {NoStop}%
\bibitem [{\citenamefont {Hellsvik}\ \emph {et~al.}(2019)\citenamefont
  {Hellsvik}, \citenamefont {Thonig}, \citenamefont {Modin}, \citenamefont
  {Iu\ifmmode~\mbox{\c{s}}\else \c{s}\fi{}an}, \citenamefont {Bergman},
  \citenamefont {Eriksson}, \citenamefont {Bergqvist},\ and\ \citenamefont
  {Delin}}]{Hellsvik2019}%
  \BibitemOpen
  \bibfield  {author} {\bibinfo {author} {\bibfnamefont {J.}~\bibnamefont
  {Hellsvik}}, \bibinfo {author} {\bibfnamefont {D.}~\bibnamefont {Thonig}},
  \bibinfo {author} {\bibfnamefont {K.}~\bibnamefont {Modin}}, \bibinfo
  {author} {\bibfnamefont {D.}~\bibnamefont {Iu\ifmmode~\mbox{\c{s}}\else
  \c{s}\fi{}an}}, \bibinfo {author} {\bibfnamefont {A.}~\bibnamefont
  {Bergman}}, \bibinfo {author} {\bibfnamefont {O.}~\bibnamefont {Eriksson}},
  \bibinfo {author} {\bibfnamefont {L.}~\bibnamefont {Bergqvist}},\ and\
  \bibinfo {author} {\bibfnamefont {A.}~\bibnamefont {Delin}},\ }\bibfield
  {title} {\bibinfo {title} {{General method for atomistic spin-lattice
  dynamics with first-principles accuracy}},\ }\href
  {https://doi.org/10.1103/PhysRevB.99.104302} {\bibfield  {journal} {\bibinfo
  {journal} {Phys. Rev. B}\ }\textbf {\bibinfo {volume} {99}},\ \bibinfo
  {pages} {104302} (\bibinfo {year} {2019})}\BibitemShut {NoStop}%
\bibitem [{\citenamefont {Mankovsky}\ \emph {et~al.}(2022)\citenamefont
  {Mankovsky}, \citenamefont {Polesya}, \citenamefont {Lange}, \citenamefont
  {Wei\ss{}enhofer}, \citenamefont {Nowak},\ and\ \citenamefont
  {Ebert}}]{Mankovsky2022}%
  \BibitemOpen
  \bibfield  {author} {\bibinfo {author} {\bibfnamefont {S.}~\bibnamefont
  {Mankovsky}}, \bibinfo {author} {\bibfnamefont {S.}~\bibnamefont {Polesya}},
  \bibinfo {author} {\bibfnamefont {H.}~\bibnamefont {Lange}}, \bibinfo
  {author} {\bibfnamefont {M.}~\bibnamefont {Wei\ss{}enhofer}}, \bibinfo
  {author} {\bibfnamefont {U.}~\bibnamefont {Nowak}},\ and\ \bibinfo {author}
  {\bibfnamefont {H.}~\bibnamefont {Ebert}},\ }\bibfield  {title} {\bibinfo
  {title} {{Angular Momentum Transfer via Relativistic Spin-Lattice Coupling
  from First Principles}},\ }\href
  {https://doi.org/10.1103/PhysRevLett.129.067202} {\bibfield  {journal}
  {\bibinfo  {journal} {Phys. Rev. Lett.}\ }\textbf {\bibinfo {volume} {129}},\
  \bibinfo {pages} {067202} (\bibinfo {year} {2022})}\BibitemShut {NoStop}%
\bibitem [{SM()}]{SM}%
  \BibitemOpen
  \href@noop {} {}\bibinfo {note} {See Supplemental Material for the derivation
  of the phonon angular momentum and the magnon-phonon Hamiltonian, the bare
  bandstructures, the results for phonon angular momentum and chirality in
  other planes of the BZ, a summary of Colpa's method, and a proof that ${\bm
  L}(T\rightarrow \infty)=0$.}\BibitemShut {Stop}%
\bibitem [{\citenamefont {Coh}(2023)}]{Coh2023}%
  \BibitemOpen
  \bibfield  {author} {\bibinfo {author} {\bibfnamefont {S.}~\bibnamefont
  {Coh}},\ }\bibfield  {title} {\bibinfo {title} {{Classification of materials
  with phonon angular momentum and microscopic origin of angular momentum}},\
  }\href {https://doi.org/10.1103/PhysRevB.108.134307} {\bibfield  {journal}
  {\bibinfo  {journal} {Phys. Rev. B}\ }\textbf {\bibinfo {volume} {108}},\
  \bibinfo {pages} {134307} (\bibinfo {year} {2023})}\BibitemShut {NoStop}%
\bibitem [{\citenamefont {Ren}\ \emph {et~al.}(2024)\citenamefont {Ren},
  \citenamefont {Bonini}, \citenamefont {Stengel}, \citenamefont {Dreyer},\
  and\ \citenamefont {Vanderbilt}}]{Ren2024}%
  \BibitemOpen
  \bibfield  {author} {\bibinfo {author} {\bibfnamefont {S.}~\bibnamefont
  {Ren}}, \bibinfo {author} {\bibfnamefont {J.}~\bibnamefont {Bonini}},
  \bibinfo {author} {\bibfnamefont {M.}~\bibnamefont {Stengel}}, \bibinfo
  {author} {\bibfnamefont {C.~E.}\ \bibnamefont {Dreyer}},\ and\ \bibinfo
  {author} {\bibfnamefont {D.}~\bibnamefont {Vanderbilt}},\ }\bibfield  {title}
  {\bibinfo {title} {{Adiabatic Dynamics of Coupled Spins and Phonons in
  Magnetic Insulators}},\ }\href {https://doi.org/10.1103/PhysRevX.14.011041}
  {\bibfield  {journal} {\bibinfo  {journal} {Phys. Rev. X}\ }\textbf {\bibinfo
  {volume} {14}},\ \bibinfo {pages} {011041} (\bibinfo {year}
  {2024})}\BibitemShut {NoStop}%
\bibitem [{not({\natexlab{a}})}]{note1}%
  \BibitemOpen
  \href@noop {} {} ({\natexlab{a}}),\ \bibinfo {note} {{We use the term
  \textit{bare} modes to describe phonon or magnon modes in the absence of
  spin-lattice coupling.}}\BibitemShut {Stop}%
\bibitem [{\citenamefont {Nowak}(2007)}]{Nowak2007}%
  \BibitemOpen
  \bibfield  {author} {\bibinfo {author} {\bibfnamefont {U.}~\bibnamefont
  {Nowak}},\ }\bibinfo {title} {Classical spin models},\ in\ \href
  {https://doi.org/https://doi.org/10.1002/9780470022184.hmm205} {\emph
  {\bibinfo {booktitle} {Handbook of Magnetism and Advanced Magnetic
  Materials}}},\ \bibinfo {editor} {edited by\ \bibinfo {editor} {\bibfnamefont
  {H.}~\bibnamefont {Kronm{\"u}ller}}\ and\ \bibinfo {editor} {\bibfnamefont
  {S.}~\bibnamefont {Parkin}}}\ (\bibinfo  {publisher} {John Wiley \& Sons,
  Ltd},\ \bibinfo {year} {2007})\BibitemShut {NoStop}%
\bibitem [{\citenamefont {Lange}\ \emph {et~al.}(2023)\citenamefont {Lange},
  \citenamefont {Mankovsky}, \citenamefont {Polesya}, \citenamefont
  {Wei\ss{}enhofer}, \citenamefont {Nowak},\ and\ \citenamefont
  {Ebert}}]{Lange2023}%
  \BibitemOpen
  \bibfield  {author} {\bibinfo {author} {\bibfnamefont {H.}~\bibnamefont
  {Lange}}, \bibinfo {author} {\bibfnamefont {S.}~\bibnamefont {Mankovsky}},
  \bibinfo {author} {\bibfnamefont {S.}~\bibnamefont {Polesya}}, \bibinfo
  {author} {\bibfnamefont {M.}~\bibnamefont {Wei\ss{}enhofer}}, \bibinfo
  {author} {\bibfnamefont {U.}~\bibnamefont {Nowak}},\ and\ \bibinfo {author}
  {\bibfnamefont {H.}~\bibnamefont {Ebert}},\ }\bibfield  {title} {\bibinfo
  {title} {{Calculating spin-lattice interactions in ferro- and
  antiferromagnets: The role of symmetry, dimension, and frustration}},\ }\href
  {https://doi.org/10.1103/PhysRevB.107.115176} {\bibfield  {journal} {\bibinfo
   {journal} {Phys. Rev. B}\ }\textbf {\bibinfo {volume} {107}},\ \bibinfo
  {pages} {115176} (\bibinfo {year} {2023})}\BibitemShut {NoStop}%
\bibitem [{\citenamefont {Mankovsky}\ \emph {et~al.}(2023)\citenamefont
  {Mankovsky}, \citenamefont {Lange}, \citenamefont {Polesya},\ and\
  \citenamefont {Ebert}}]{Mankovsky2023}%
  \BibitemOpen
  \bibfield  {author} {\bibinfo {author} {\bibfnamefont {S.}~\bibnamefont
  {Mankovsky}}, \bibinfo {author} {\bibfnamefont {H.}~\bibnamefont {Lange}},
  \bibinfo {author} {\bibfnamefont {S.}~\bibnamefont {Polesya}},\ and\ \bibinfo
  {author} {\bibfnamefont {H.}~\bibnamefont {Ebert}},\ }\bibfield  {title}
  {\bibinfo {title} {{Spin-lattice interaction parameters from first
  principles: Theory and implementation}},\ }\href
  {https://doi.org/10.1103/PhysRevB.107.144428} {\bibfield  {journal} {\bibinfo
   {journal} {Phys. Rev. B}\ }\textbf {\bibinfo {volume} {107}},\ \bibinfo
  {pages} {144428} (\bibinfo {year} {2023})}\BibitemShut {NoStop}%
\bibitem [{\citenamefont {Miranda}\ \emph {et~al.}(2024)\citenamefont
  {Miranda}, \citenamefont {Pankratova}, \citenamefont {Weißenhofer},
  \citenamefont {Klautau}, \citenamefont {Thonig}, \citenamefont {Pereiro},
  \citenamefont {Sjöqvist}, \citenamefont {Delin}, \citenamefont {Katsnelson},
  \citenamefont {Eriksson},\ and\ \citenamefont {Bergman}}]{Miranda2024}%
  \BibitemOpen
  \bibfield  {author} {\bibinfo {author} {\bibfnamefont {I.~P.}\ \bibnamefont
  {Miranda}}, \bibinfo {author} {\bibfnamefont {M.}~\bibnamefont {Pankratova}},
  \bibinfo {author} {\bibfnamefont {M.}~\bibnamefont {Weißenhofer}}, \bibinfo
  {author} {\bibfnamefont {A.~B.}\ \bibnamefont {Klautau}}, \bibinfo {author}
  {\bibfnamefont {D.}~\bibnamefont {Thonig}}, \bibinfo {author} {\bibfnamefont
  {M.}~\bibnamefont {Pereiro}}, \bibinfo {author} {\bibfnamefont
  {E.}~\bibnamefont {Sjöqvist}}, \bibinfo {author} {\bibfnamefont
  {A.}~\bibnamefont {Delin}}, \bibinfo {author} {\bibfnamefont {M.~I.}\
  \bibnamefont {Katsnelson}}, \bibinfo {author} {\bibfnamefont
  {O.}~\bibnamefont {Eriksson}},\ and\ \bibinfo {author} {\bibfnamefont
  {A.}~\bibnamefont {Bergman}},\ }\href {https://arxiv.org/abs/2409.18274}
  {\bibinfo {title} {{Spin-lattice couplings in $3d$ ferromagnets: analysis
  from first-principles}}} (\bibinfo {year} {2024}),\ \Eprint
  {https://arxiv.org/abs/2409.18274} {arXiv:2409.18274 [cond-mat.mtrl-sci]}
  \BibitemShut {NoStop}%
\bibitem [{\citenamefont {Kahana}\ \emph {et~al.}(2024)\citenamefont {Kahana},
  \citenamefont {Lopez},\ and\ \citenamefont {Juraschek}}]{Kahana2024}%
  \BibitemOpen
  \bibfield  {author} {\bibinfo {author} {\bibfnamefont {T.}~\bibnamefont
  {Kahana}}, \bibinfo {author} {\bibfnamefont {D.~A.~B.}\ \bibnamefont
  {Lopez}},\ and\ \bibinfo {author} {\bibfnamefont {D.~M.}\ \bibnamefont
  {Juraschek}},\ }\bibfield  {title} {\bibinfo {title} {{Light-induced
  magnetization from magnonic rectification}},\ }\href
  {https://doi.org/10.1126/sciadv.ado0722} {\bibfield  {journal} {\bibinfo
  {journal} {Sci. Adv.}\ }\textbf {\bibinfo {volume} {10}},\ \bibinfo {pages}
  {eado0722} (\bibinfo {year} {2024})}\BibitemShut {NoStop}%
\bibitem [{\citenamefont {Mead}\ and\ \citenamefont
  {Truhlar}(1979)}]{MeadTruhlar1979}%
  \BibitemOpen
  \bibfield  {author} {\bibinfo {author} {\bibfnamefont {C.~A.}\ \bibnamefont
  {Mead}}\ and\ \bibinfo {author} {\bibfnamefont {D.~G.}\ \bibnamefont
  {Truhlar}},\ }\bibfield  {title} {\bibinfo {title} {{{On the determination of
  Born–Oppenheimer nuclear motion wave functions including complications due
  to conical intersections and identical nuclei}}},\ }\href
  {https://doi.org/10.1063/1.437734} {\bibfield  {journal} {\bibinfo  {journal}
  {J. Chem. Phys.}\ }\textbf {\bibinfo {volume} {70}},\ \bibinfo {pages} {2284}
  (\bibinfo {year} {1979})}\BibitemShut {NoStop}%
\bibitem [{\citenamefont {Born}\ and\ \citenamefont
  {Huang}(1996)}]{BornHuang1996}%
  \BibitemOpen
  \bibfield  {author} {\bibinfo {author} {\bibfnamefont {M.}~\bibnamefont
  {Born}}\ and\ \bibinfo {author} {\bibfnamefont {K.}~\bibnamefont {Huang}},\
  }\href {https://doi.org/10.1093/oso/9780192670083.001.0001} {\emph {\bibinfo
  {title} {{{Dynamical Theory Of Crystal Lattices}}}}}\ (\bibinfo  {publisher}
  {Oxford University Press},\ \bibinfo {year} {1996})\BibitemShut {NoStop}%
\bibitem [{\citenamefont {Zhang}\ \emph {et~al.}(2011)\citenamefont {Zhang},
  \citenamefont {Ren}, \citenamefont {Wang},\ and\ \citenamefont
  {Li}}]{Zhang2011}%
  \BibitemOpen
  \bibfield  {author} {\bibinfo {author} {\bibfnamefont {L.}~\bibnamefont
  {Zhang}}, \bibinfo {author} {\bibfnamefont {J.}~\bibnamefont {Ren}}, \bibinfo
  {author} {\bibfnamefont {J.-S.}\ \bibnamefont {Wang}},\ and\ \bibinfo
  {author} {\bibfnamefont {B.}~\bibnamefont {Li}},\ }\bibfield  {title}
  {\bibinfo {title} {{The phonon {H}all effect: theory and application}},\
  }\href {https://doi.org/10.1088/0953-8984/23/30/305402} {\bibfield  {journal}
  {\bibinfo  {journal} {J. Phys.: Condens. Matter}\ }\textbf {\bibinfo {volume}
  {23}},\ \bibinfo {pages} {305402} (\bibinfo {year} {2011})}\BibitemShut
  {NoStop}%
\bibitem [{\citenamefont {Holstein}\ and\ \citenamefont
  {Primakoff}(1940)}]{Holstein1940}%
  \BibitemOpen
  \bibfield  {author} {\bibinfo {author} {\bibfnamefont {T.}~\bibnamefont
  {Holstein}}\ and\ \bibinfo {author} {\bibfnamefont {H.}~\bibnamefont
  {Primakoff}},\ }\bibfield  {title} {\bibinfo {title} {{Field Dependence of
  the Intrinsic Domain Magnetization of a Ferromagnet}},\ }\href
  {https://doi.org/10.1103/PhysRev.58.1098} {\bibfield  {journal} {\bibinfo
  {journal} {Phys. Rev.}\ }\textbf {\bibinfo {volume} {58}},\ \bibinfo {pages}
  {1098} (\bibinfo {year} {1940})}\BibitemShut {NoStop}%
\bibitem [{\citenamefont {Mryasov}\ \emph {et~al.}(1996)\citenamefont
  {Mryasov}, \citenamefont {Freeman},\ and\ \citenamefont
  {Liechtenstein}}]{Mryasov1996}%
  \BibitemOpen
  \bibfield  {author} {\bibinfo {author} {\bibfnamefont {O.~N.}\ \bibnamefont
  {Mryasov}}, \bibinfo {author} {\bibfnamefont {A.~J.}\ \bibnamefont
  {Freeman}},\ and\ \bibinfo {author} {\bibfnamefont {A.~I.}\ \bibnamefont
  {Liechtenstein}},\ }\bibfield  {title} {\bibinfo {title} {{{Theory of
  non‐Heisenberg exchange: Results for localized and itinerant magnets}}},\
  }\href {https://doi.org/10.1063/1.361678} {\bibfield  {journal} {\bibinfo
  {journal} {J. Appl. Phys.}\ }\textbf {\bibinfo {volume} {79}},\ \bibinfo
  {pages} {4805} (\bibinfo {year} {1996})}\BibitemShut {NoStop}%
\bibitem [{\citenamefont {Razdolski}\ \emph {et~al.}(2017)\citenamefont
  {Razdolski}, \citenamefont {Alekhin}, \citenamefont {Ilin}, \citenamefont
  {Meyburg}, \citenamefont {Roddatis}, \citenamefont {Diesing}, \citenamefont
  {Bovensiepen},\ and\ \citenamefont {Melnikov}}]{Razdolski2017}%
  \BibitemOpen
  \bibfield  {author} {\bibinfo {author} {\bibfnamefont {I.}~\bibnamefont
  {Razdolski}}, \bibinfo {author} {\bibfnamefont {A.}~\bibnamefont {Alekhin}},
  \bibinfo {author} {\bibfnamefont {N.}~\bibnamefont {Ilin}}, \bibinfo {author}
  {\bibfnamefont {J.~P.}\ \bibnamefont {Meyburg}}, \bibinfo {author}
  {\bibfnamefont {V.}~\bibnamefont {Roddatis}}, \bibinfo {author}
  {\bibfnamefont {D.}~\bibnamefont {Diesing}}, \bibinfo {author} {\bibfnamefont
  {U.}~\bibnamefont {Bovensiepen}},\ and\ \bibinfo {author} {\bibfnamefont
  {A.}~\bibnamefont {Melnikov}},\ }\bibfield  {title} {\bibinfo {title}
  {{Nanoscale interface confinement of ultrafast spin transfer torque driving
  non-uniform spin dynamics}},\ }\href {https://doi.org/10.1038/ncomms15007}
  {\bibfield  {journal} {\bibinfo  {journal} {Nature Commun.}\ }\textbf
  {\bibinfo {volume} {8}},\ \bibinfo {pages} {15007} (\bibinfo {year}
  {2017})}\BibitemShut {NoStop}%
\bibitem [{\citenamefont {Giannozzi}\ \emph {et~al.}(2009)\citenamefont
  {Giannozzi}, \citenamefont {Baroni}, \citenamefont {Bonini}, \citenamefont
  {Calandra}, \citenamefont {Car}, \citenamefont {Cavazzoni}, \citenamefont
  {Ceresoli}, \citenamefont {Chiarotti}, \citenamefont {Cococcioni},
  \citenamefont {Dabo}, \citenamefont {Corso}, \citenamefont {de~Gironcoli},
  \citenamefont {Fabris}, \citenamefont {Fratesi}, \citenamefont {Gebauer},
  \citenamefont {Gerstmann}, \citenamefont {Gougoussis}, \citenamefont
  {Kokalj}, \citenamefont {Lazzeri}, \citenamefont {Martin-Samos},
  \citenamefont {Marzari}, \citenamefont {Mauri}, \citenamefont {Mazzarello},
  \citenamefont {Paolini}, \citenamefont {Pasquarello}, \citenamefont
  {Paulatto}, \citenamefont {Sbraccia}, \citenamefont {Scandolo}, \citenamefont
  {Sclauzero}, \citenamefont {Seitsonen}, \citenamefont {Smogunov},
  \citenamefont {Umari},\ and\ \citenamefont {Wentzcovitch}}]{Giannozzi2009}%
  \BibitemOpen
  \bibfield  {author} {\bibinfo {author} {\bibfnamefont {P.}~\bibnamefont
  {Giannozzi}}, \bibinfo {author} {\bibfnamefont {S.}~\bibnamefont {Baroni}},
  \bibinfo {author} {\bibfnamefont {N.}~\bibnamefont {Bonini}}, \bibinfo
  {author} {\bibfnamefont {M.}~\bibnamefont {Calandra}}, \bibinfo {author}
  {\bibfnamefont {R.}~\bibnamefont {Car}}, \bibinfo {author} {\bibfnamefont
  {C.}~\bibnamefont {Cavazzoni}}, \bibinfo {author} {\bibfnamefont
  {D.}~\bibnamefont {Ceresoli}}, \bibinfo {author} {\bibfnamefont {G.~L.}\
  \bibnamefont {Chiarotti}}, \bibinfo {author} {\bibfnamefont {M.}~\bibnamefont
  {Cococcioni}}, \bibinfo {author} {\bibfnamefont {I.}~\bibnamefont {Dabo}},
  \bibinfo {author} {\bibfnamefont {A.~D.}\ \bibnamefont {Corso}}, \bibinfo
  {author} {\bibfnamefont {S.}~\bibnamefont {de~Gironcoli}}, \bibinfo {author}
  {\bibfnamefont {S.}~\bibnamefont {Fabris}}, \bibinfo {author} {\bibfnamefont
  {G.}~\bibnamefont {Fratesi}}, \bibinfo {author} {\bibfnamefont
  {R.}~\bibnamefont {Gebauer}}, \bibinfo {author} {\bibfnamefont
  {U.}~\bibnamefont {Gerstmann}}, \bibinfo {author} {\bibfnamefont
  {C.}~\bibnamefont {Gougoussis}}, \bibinfo {author} {\bibfnamefont
  {A.}~\bibnamefont {Kokalj}}, \bibinfo {author} {\bibfnamefont
  {M.}~\bibnamefont {Lazzeri}}, \bibinfo {author} {\bibfnamefont
  {L.}~\bibnamefont {Martin-Samos}}, \bibinfo {author} {\bibfnamefont
  {N.}~\bibnamefont {Marzari}}, \bibinfo {author} {\bibfnamefont
  {F.}~\bibnamefont {Mauri}}, \bibinfo {author} {\bibfnamefont
  {R.}~\bibnamefont {Mazzarello}}, \bibinfo {author} {\bibfnamefont
  {S.}~\bibnamefont {Paolini}}, \bibinfo {author} {\bibfnamefont
  {A.}~\bibnamefont {Pasquarello}}, \bibinfo {author} {\bibfnamefont
  {L.}~\bibnamefont {Paulatto}}, \bibinfo {author} {\bibfnamefont
  {C.}~\bibnamefont {Sbraccia}}, \bibinfo {author} {\bibfnamefont
  {S.}~\bibnamefont {Scandolo}}, \bibinfo {author} {\bibfnamefont
  {G.}~\bibnamefont {Sclauzero}}, \bibinfo {author} {\bibfnamefont {A.~P.}\
  \bibnamefont {Seitsonen}}, \bibinfo {author} {\bibfnamefont {A.}~\bibnamefont
  {Smogunov}}, \bibinfo {author} {\bibfnamefont {P.}~\bibnamefont {Umari}},\
  and\ \bibinfo {author} {\bibfnamefont {R.~M.}\ \bibnamefont {Wentzcovitch}},\
  }\bibfield  {title} {\bibinfo {title} {{QUANTUM ESPRESSO: a modular and
  open-source software project for quantum simulations of materials}},\ }\href
  {https://doi.org/10.1088/0953-8984/21/39/395502} {\bibfield  {journal}
  {\bibinfo  {journal} {J. Phys.: Condens. Matter}\ }\textbf {\bibinfo {volume}
  {21}},\ \bibinfo {pages} {395502} (\bibinfo {year} {2009})}\BibitemShut
  {NoStop}%
\bibitem [{\citenamefont {Giannozzi}\ \emph {et~al.}(2017)\citenamefont
  {Giannozzi}, \citenamefont {Andreussi}, \citenamefont {Brumme}, \citenamefont
  {Bunau}, \citenamefont {Nardelli}, \citenamefont {Calandra}, \citenamefont
  {Car}, \citenamefont {Cavazzoni}, \citenamefont {Ceresoli}, \citenamefont
  {Cococcioni}, \citenamefont {Colonna}, \citenamefont {Carnimeo},
  \citenamefont {Corso}, \citenamefont {de~Gironcoli}, \citenamefont {Delugas},
  \citenamefont {DiStasio}, \citenamefont {Ferretti}, \citenamefont {Floris},
  \citenamefont {Fratesi}, \citenamefont {Fugallo}, \citenamefont {Gebauer},
  \citenamefont {Gerstmann}, \citenamefont {Giustino}, \citenamefont {Gorni},
  \citenamefont {Jia}, \citenamefont {Kawamura}, \citenamefont {Ko},
  \citenamefont {Kokalj}, \citenamefont {Küçükbenli}, \citenamefont
  {Lazzeri}, \citenamefont {Marsili}, \citenamefont {Marzari}, \citenamefont
  {Mauri}, \citenamefont {Nguyen}, \citenamefont {Nguyen}, \citenamefont {de-la
  Roza}, \citenamefont {Paulatto}, \citenamefont {Poncé}, \citenamefont
  {Rocca}, \citenamefont {Sabatini}, \citenamefont {Santra}, \citenamefont
  {Schlipf}, \citenamefont {Seitsonen}, \citenamefont {Smogunov}, \citenamefont
  {Timrov}, \citenamefont {Thonhauser}, \citenamefont {Umari}, \citenamefont
  {Vast}, \citenamefont {Wu},\ and\ \citenamefont {Baroni}}]{Giannozzi2017}%
  \BibitemOpen
  \bibfield  {author} {\bibinfo {author} {\bibfnamefont {P.}~\bibnamefont
  {Giannozzi}}, \bibinfo {author} {\bibfnamefont {O.}~\bibnamefont
  {Andreussi}}, \bibinfo {author} {\bibfnamefont {T.}~\bibnamefont {Brumme}},
  \bibinfo {author} {\bibfnamefont {O.}~\bibnamefont {Bunau}}, \bibinfo
  {author} {\bibfnamefont {M.~B.}\ \bibnamefont {Nardelli}}, \bibinfo {author}
  {\bibfnamefont {M.}~\bibnamefont {Calandra}}, \bibinfo {author}
  {\bibfnamefont {R.}~\bibnamefont {Car}}, \bibinfo {author} {\bibfnamefont
  {C.}~\bibnamefont {Cavazzoni}}, \bibinfo {author} {\bibfnamefont
  {D.}~\bibnamefont {Ceresoli}}, \bibinfo {author} {\bibfnamefont
  {M.}~\bibnamefont {Cococcioni}}, \bibinfo {author} {\bibfnamefont
  {N.}~\bibnamefont {Colonna}}, \bibinfo {author} {\bibfnamefont
  {I.}~\bibnamefont {Carnimeo}}, \bibinfo {author} {\bibfnamefont {A.~D.}\
  \bibnamefont {Corso}}, \bibinfo {author} {\bibfnamefont {S.}~\bibnamefont
  {de~Gironcoli}}, \bibinfo {author} {\bibfnamefont {P.}~\bibnamefont
  {Delugas}}, \bibinfo {author} {\bibfnamefont {R.~A.}\ \bibnamefont
  {DiStasio}}, \bibinfo {author} {\bibfnamefont {A.}~\bibnamefont {Ferretti}},
  \bibinfo {author} {\bibfnamefont {A.}~\bibnamefont {Floris}}, \bibinfo
  {author} {\bibfnamefont {G.}~\bibnamefont {Fratesi}}, \bibinfo {author}
  {\bibfnamefont {G.}~\bibnamefont {Fugallo}}, \bibinfo {author} {\bibfnamefont
  {R.}~\bibnamefont {Gebauer}}, \bibinfo {author} {\bibfnamefont
  {U.}~\bibnamefont {Gerstmann}}, \bibinfo {author} {\bibfnamefont
  {F.}~\bibnamefont {Giustino}}, \bibinfo {author} {\bibfnamefont
  {T.}~\bibnamefont {Gorni}}, \bibinfo {author} {\bibfnamefont
  {J.}~\bibnamefont {Jia}}, \bibinfo {author} {\bibfnamefont {M.}~\bibnamefont
  {Kawamura}}, \bibinfo {author} {\bibfnamefont {H.-Y.}\ \bibnamefont {Ko}},
  \bibinfo {author} {\bibfnamefont {A.}~\bibnamefont {Kokalj}}, \bibinfo
  {author} {\bibfnamefont {E.}~\bibnamefont {Küçükbenli}}, \bibinfo {author}
  {\bibfnamefont {M.}~\bibnamefont {Lazzeri}}, \bibinfo {author} {\bibfnamefont
  {M.}~\bibnamefont {Marsili}}, \bibinfo {author} {\bibfnamefont
  {N.}~\bibnamefont {Marzari}}, \bibinfo {author} {\bibfnamefont
  {F.}~\bibnamefont {Mauri}}, \bibinfo {author} {\bibfnamefont {N.~L.}\
  \bibnamefont {Nguyen}}, \bibinfo {author} {\bibfnamefont {H.-V.}\
  \bibnamefont {Nguyen}}, \bibinfo {author} {\bibfnamefont {A.~O.}\
  \bibnamefont {de-la Roza}}, \bibinfo {author} {\bibfnamefont
  {L.}~\bibnamefont {Paulatto}}, \bibinfo {author} {\bibfnamefont
  {S.}~\bibnamefont {Poncé}}, \bibinfo {author} {\bibfnamefont
  {D.}~\bibnamefont {Rocca}}, \bibinfo {author} {\bibfnamefont
  {R.}~\bibnamefont {Sabatini}}, \bibinfo {author} {\bibfnamefont
  {B.}~\bibnamefont {Santra}}, \bibinfo {author} {\bibfnamefont
  {M.}~\bibnamefont {Schlipf}}, \bibinfo {author} {\bibfnamefont {A.~P.}\
  \bibnamefont {Seitsonen}}, \bibinfo {author} {\bibfnamefont {A.}~\bibnamefont
  {Smogunov}}, \bibinfo {author} {\bibfnamefont {I.}~\bibnamefont {Timrov}},
  \bibinfo {author} {\bibfnamefont {T.}~\bibnamefont {Thonhauser}}, \bibinfo
  {author} {\bibfnamefont {P.}~\bibnamefont {Umari}}, \bibinfo {author}
  {\bibfnamefont {N.}~\bibnamefont {Vast}}, \bibinfo {author} {\bibfnamefont
  {X.}~\bibnamefont {Wu}},\ and\ \bibinfo {author} {\bibfnamefont
  {S.}~\bibnamefont {Baroni}},\ }\bibfield  {title} {\bibinfo {title}
  {{Advanced capabilities for materials modelling with Quantum ESPRESSO}},\
  }\href {https://doi.org/10.1088/1361-648X/aa8f79} {\bibfield  {journal}
  {\bibinfo  {journal} {J. Phys.: Condens. Matter}\ }\textbf {\bibinfo {volume}
  {29}},\ \bibinfo {pages} {465901} (\bibinfo {year} {2017})}\BibitemShut
  {NoStop}%
\bibitem [{\citenamefont {Colpa}(1978)}]{Colpa1978}%
  \BibitemOpen
  \bibfield  {author} {\bibinfo {author} {\bibfnamefont {J.}~\bibnamefont
  {Colpa}},\ }\bibfield  {title} {\bibinfo {title} {{Diagonalization of the
  quadratic boson hamiltonian}},\ }\href
  {https://doi.org/https://doi.org/10.1016/0378-4371(78)90160-7} {\bibfield
  {journal} {\bibinfo  {journal} {Physica A: Statistical Mechanics and its
  Applications}\ }\textbf {\bibinfo {volume} {93}},\ \bibinfo {pages} {327}
  (\bibinfo {year} {1978})}\BibitemShut {NoStop}%
\bibitem [{\citenamefont {Li}\ \emph {et~al.}(2021)\citenamefont {Li},
  \citenamefont {Zhao}, \citenamefont {Zhang}, \citenamefont {Hoffmann},\ and\
  \citenamefont {Novosad}}]{Li2021}%
  \BibitemOpen
  \bibfield  {author} {\bibinfo {author} {\bibfnamefont {Y.}~\bibnamefont
  {Li}}, \bibinfo {author} {\bibfnamefont {C.}~\bibnamefont {Zhao}}, \bibinfo
  {author} {\bibfnamefont {W.}~\bibnamefont {Zhang}}, \bibinfo {author}
  {\bibfnamefont {A.}~\bibnamefont {Hoffmann}},\ and\ \bibinfo {author}
  {\bibfnamefont {V.}~\bibnamefont {Novosad}},\ }\bibfield  {title} {\bibinfo
  {title} {{{Advances in coherent coupling between magnons and acoustic
  phonons}}},\ }\href {https://doi.org/10.1063/5.0047054} {\bibfield  {journal}
  {\bibinfo  {journal} {APL Materials}\ }\textbf {\bibinfo {volume} {9}},\
  \bibinfo {pages} {060902} (\bibinfo {year} {2021})}\BibitemShut {NoStop}%
\bibitem [{\citenamefont {Kittel}(1949)}]{Kittel1949}%
  \BibitemOpen
  \bibfield  {author} {\bibinfo {author} {\bibfnamefont {C.}~\bibnamefont
  {Kittel}},\ }\bibfield  {title} {\bibinfo {title} {{Physical Theory of
  Ferromagnetic Domains}},\ }\href {https://doi.org/10.1103/RevModPhys.21.541}
  {\bibfield  {journal} {\bibinfo  {journal} {Rev. Mod. Phys.}\ }\textbf
  {\bibinfo {volume} {21}},\ \bibinfo {pages} {541} (\bibinfo {year}
  {1949})}\BibitemShut {NoStop}%
\bibitem [{\citenamefont {Wei\ss{}enhofer}\ \emph {et~al.}(2023)\citenamefont
  {Wei\ss{}enhofer}, \citenamefont {Lange}, \citenamefont {Kamra},
  \citenamefont {Mankovsky}, \citenamefont {Polesya}, \citenamefont {Ebert},\
  and\ \citenamefont {Nowak}}]{Weissenhofer2023}%
  \BibitemOpen
  \bibfield  {author} {\bibinfo {author} {\bibfnamefont {M.}~\bibnamefont
  {Wei\ss{}enhofer}}, \bibinfo {author} {\bibfnamefont {H.}~\bibnamefont
  {Lange}}, \bibinfo {author} {\bibfnamefont {A.}~\bibnamefont {Kamra}},
  \bibinfo {author} {\bibfnamefont {S.}~\bibnamefont {Mankovsky}}, \bibinfo
  {author} {\bibfnamefont {S.}~\bibnamefont {Polesya}}, \bibinfo {author}
  {\bibfnamefont {H.}~\bibnamefont {Ebert}},\ and\ \bibinfo {author}
  {\bibfnamefont {U.}~\bibnamefont {Nowak}},\ }\bibfield  {title} {\bibinfo
  {title} {{Rotationally invariant formulation of spin-lattice coupling in
  multiscale modeling}},\ }\href {https://doi.org/10.1103/PhysRevB.108.L060404}
  {\bibfield  {journal} {\bibinfo  {journal} {Phys. Rev. B}\ }\textbf {\bibinfo
  {volume} {108}},\ \bibinfo {pages} {L060404} (\bibinfo {year}
  {2023})}\BibitemShut {NoStop}%
\bibitem [{\citenamefont {White}\ \emph {et~al.}(1965)\citenamefont {White},
  \citenamefont {Sparks},\ and\ \citenamefont {Ortenburger}}]{White1965}%
  \BibitemOpen
  \bibfield  {author} {\bibinfo {author} {\bibfnamefont {R.~M.}\ \bibnamefont
  {White}}, \bibinfo {author} {\bibfnamefont {M.}~\bibnamefont {Sparks}},\ and\
  \bibinfo {author} {\bibfnamefont {I.}~\bibnamefont {Ortenburger}},\
  }\bibfield  {title} {\bibinfo {title} {{Diagonalization of the
  Antiferromagnetic Magnon-Phonon Interaction}},\ }\href
  {https://doi.org/10.1103/PhysRev.139.A450} {\bibfield  {journal} {\bibinfo
  {journal} {Phys. Rev.}\ }\textbf {\bibinfo {volume} {139}},\ \bibinfo {pages}
  {A450} (\bibinfo {year} {1965})}\BibitemShut {NoStop}%
\bibitem [{\citenamefont {R\"uckriegel}\ \emph {et~al.}(2014)\citenamefont
  {R\"uckriegel}, \citenamefont {Kopietz}, \citenamefont {Bozhko},
  \citenamefont {Serga},\ and\ \citenamefont {Hillebrands}}]{Rueckriegel2014}%
  \BibitemOpen
  \bibfield  {author} {\bibinfo {author} {\bibfnamefont {A.}~\bibnamefont
  {R\"uckriegel}}, \bibinfo {author} {\bibfnamefont {P.}~\bibnamefont
  {Kopietz}}, \bibinfo {author} {\bibfnamefont {D.~A.}\ \bibnamefont {Bozhko}},
  \bibinfo {author} {\bibfnamefont {A.~A.}\ \bibnamefont {Serga}},\ and\
  \bibinfo {author} {\bibfnamefont {B.}~\bibnamefont {Hillebrands}},\
  }\bibfield  {title} {\bibinfo {title} {Magnetoelastic modes and lifetime of
  magnons in thin yttrium iron garnet films},\ }\href
  {https://doi.org/10.1103/PhysRevB.89.184413} {\bibfield  {journal} {\bibinfo
  {journal} {Phys. Rev. B}\ }\textbf {\bibinfo {volume} {89}},\ \bibinfo
  {pages} {184413} (\bibinfo {year} {2014})}\BibitemShut {NoStop}%
\bibitem [{\citenamefont {Kamra}\ \emph {et~al.}(2015)\citenamefont {Kamra},
  \citenamefont {Keshtgar}, \citenamefont {Yan},\ and\ \citenamefont
  {Bauer}}]{Kamra2015}%
  \BibitemOpen
  \bibfield  {author} {\bibinfo {author} {\bibfnamefont {A.}~\bibnamefont
  {Kamra}}, \bibinfo {author} {\bibfnamefont {H.}~\bibnamefont {Keshtgar}},
  \bibinfo {author} {\bibfnamefont {P.}~\bibnamefont {Yan}},\ and\ \bibinfo
  {author} {\bibfnamefont {G.~E.~W.}\ \bibnamefont {Bauer}},\ }\bibfield
  {title} {\bibinfo {title} {Coherent elastic excitation of spin waves},\
  }\href {https://doi.org/10.1103/PhysRevB.91.104409} {\bibfield  {journal}
  {\bibinfo  {journal} {Phys. Rev. B}\ }\textbf {\bibinfo {volume} {91}},\
  \bibinfo {pages} {104409} (\bibinfo {year} {2015})}\BibitemShut {NoStop}%
\bibitem [{\citenamefont {Streib}\ \emph {et~al.}(2019)\citenamefont {Streib},
  \citenamefont {Vidal-Silva}, \citenamefont {Shen},\ and\ \citenamefont
  {Bauer}}]{Streib2019}%
  \BibitemOpen
  \bibfield  {author} {\bibinfo {author} {\bibfnamefont {S.}~\bibnamefont
  {Streib}}, \bibinfo {author} {\bibfnamefont {N.}~\bibnamefont {Vidal-Silva}},
  \bibinfo {author} {\bibfnamefont {K.}~\bibnamefont {Shen}},\ and\ \bibinfo
  {author} {\bibfnamefont {G.~E.~W.}\ \bibnamefont {Bauer}},\ }\bibfield
  {title} {\bibinfo {title} {{Magnon-phonon interactions in magnetic
  insulators}},\ }\href {https://doi.org/10.1103/PhysRevB.99.184442} {\bibfield
   {journal} {\bibinfo  {journal} {Phys. Rev. B}\ }\textbf {\bibinfo {volume}
  {99}},\ \bibinfo {pages} {184442} (\bibinfo {year} {2019})}\BibitemShut
  {NoStop}%
\bibitem [{\citenamefont {Gurevich}\ and\ \citenamefont
  {Melkov}(2020)}]{Gurevich2020}%
  \BibitemOpen
  \bibfield  {author} {\bibinfo {author} {\bibfnamefont {A.~G.}\ \bibnamefont
  {Gurevich}}\ and\ \bibinfo {author} {\bibfnamefont {G.~A.}\ \bibnamefont
  {Melkov}},\ }\href
  {https://www.taylorfrancis.com/books/mono/10.1201/9780138748487/magnetization-oscillations-waves-melkov-gurevich}
  {\emph {\bibinfo {title} {{Magnetization oscillations and waves}}}}\
  (\bibinfo  {publisher} {CRC press},\ \bibinfo {year} {2020})\BibitemShut
  {NoStop}%
\bibitem [{not({\natexlab{b}})}]{note3}%
  \BibitemOpen
  \href@noop {} {} ({\natexlab{b}}),\ \bibinfo {note} {{Classically, a spin
  $\vec{S}$ in a magnetic field $\vec{H}$ precesses as
  $\dot{\vec{S}}=-|\gamma/\mu_\mathrm{s}| (\vec{S} \times \vec{H})$, i.e., it
  follows a counterclockwise motion.}}\BibitemShut {Stop}%
\bibitem [{\citenamefont {Cui}\ \emph {et~al.}(2023)\citenamefont {Cui},
  \citenamefont {Bostr{\"o}m}, \citenamefont {Ozerov}, \citenamefont {Wu},
  \citenamefont {Jiang}, \citenamefont {Chu}, \citenamefont {Li}, \citenamefont
  {Liu}, \citenamefont {Xu}, \citenamefont {Rubio},\ and\ \citenamefont
  {Zhang}}]{Cui2023}%
  \BibitemOpen
  \bibfield  {author} {\bibinfo {author} {\bibfnamefont {J.}~\bibnamefont
  {Cui}}, \bibinfo {author} {\bibfnamefont {E.~V.}\ \bibnamefont
  {Bostr{\"o}m}}, \bibinfo {author} {\bibfnamefont {M.}~\bibnamefont {Ozerov}},
  \bibinfo {author} {\bibfnamefont {F.}~\bibnamefont {Wu}}, \bibinfo {author}
  {\bibfnamefont {Q.}~\bibnamefont {Jiang}}, \bibinfo {author} {\bibfnamefont
  {J.-H.}\ \bibnamefont {Chu}}, \bibinfo {author} {\bibfnamefont
  {C.}~\bibnamefont {Li}}, \bibinfo {author} {\bibfnamefont {F.}~\bibnamefont
  {Liu}}, \bibinfo {author} {\bibfnamefont {X.}~\bibnamefont {Xu}}, \bibinfo
  {author} {\bibfnamefont {A.}~\bibnamefont {Rubio}},\ and\ \bibinfo {author}
  {\bibfnamefont {Q.}~\bibnamefont {Zhang}},\ }\bibfield  {title} {\bibinfo
  {title} {{Chirality selective magnon-phonon hybridization and magnon-induced
  chiral phonons in a layered zigzag antiferromagnet}},\ }\href
  {https://doi.org/10.1038/s41467-023-39123-y} {\bibfield  {journal} {\bibinfo
  {journal} {Nature Commun.}\ }\textbf {\bibinfo {volume} {14}},\ \bibinfo
  {pages} {3396} (\bibinfo {year} {2023})}\BibitemShut {NoStop}%
\bibitem [{\citenamefont {Dresselhaus}\ \emph {et~al.}(2007)\citenamefont
  {Dresselhaus}, \citenamefont {Dresselhaus},\ and\ \citenamefont
  {Jorio}}]{Dresselhaus2007}%
  \BibitemOpen
  \bibfield  {author} {\bibinfo {author} {\bibfnamefont {M.~S.}\ \bibnamefont
  {Dresselhaus}}, \bibinfo {author} {\bibfnamefont {G.}~\bibnamefont
  {Dresselhaus}},\ and\ \bibinfo {author} {\bibfnamefont {A.}~\bibnamefont
  {Jorio}},\ }\href@noop {} {\emph {\bibinfo {title} {{Group theory:
  application to the physics of condensed matter}}}}\ (\bibinfo  {publisher}
  {Springer Science \& Business Media},\ \bibinfo {year} {2007})\BibitemShut
  {NoStop}%
\bibitem [{\citenamefont {Evans}\ \emph {et~al.}(2014)\citenamefont {Evans},
  \citenamefont {Fan}, \citenamefont {Chureemart}, \citenamefont {Ostler},
  \citenamefont {Ellis},\ and\ \citenamefont {Chantrell}}]{Evans2014}%
  \BibitemOpen
  \bibfield  {author} {\bibinfo {author} {\bibfnamefont {R.~F.~L.}\
  \bibnamefont {Evans}}, \bibinfo {author} {\bibfnamefont {W.~J.}\ \bibnamefont
  {Fan}}, \bibinfo {author} {\bibfnamefont {P.}~\bibnamefont {Chureemart}},
  \bibinfo {author} {\bibfnamefont {T.~A.}\ \bibnamefont {Ostler}}, \bibinfo
  {author} {\bibfnamefont {M.~O.~A.}\ \bibnamefont {Ellis}},\ and\ \bibinfo
  {author} {\bibfnamefont {R.~W.}\ \bibnamefont {Chantrell}},\ }\bibfield
  {title} {\bibinfo {title} {{Atomistic spin model simulations of magnetic
  nanomaterials}},\ }\href {https://doi.org/10.1088/0953-8984/26/10/103202}
  {\bibfield  {journal} {\bibinfo  {journal} {J. Phys.: Condens. Matter}\
  }\textbf {\bibinfo {volume} {26}},\ \bibinfo {pages} {103202} (\bibinfo
  {year} {2014})}\BibitemShut {NoStop}%
\bibitem [{not({\natexlab{c}})}]{note2}%
  \BibitemOpen
  \href@noop {} {} ({\natexlab{c}}),\ \bibinfo {note} {{A minimum of
  $B\approx\SI{4.5}{\tesla}$ is needed to shift the bare magnons energies above
  those of the bare TA phonons.}}\BibitemShut {Stop}%
\bibitem [{\citenamefont {Park}\ \emph {et~al.}(2020)\citenamefont {Park},
  \citenamefont {Nagaosa},\ and\ \citenamefont {Yang}}]{Park2020}%
  \BibitemOpen
  \bibfield  {author} {\bibinfo {author} {\bibfnamefont {S.}~\bibnamefont
  {Park}}, \bibinfo {author} {\bibfnamefont {N.}~\bibnamefont {Nagaosa}},\ and\
  \bibinfo {author} {\bibfnamefont {B.-J.}\ \bibnamefont {Yang}},\ }\bibfield
  {title} {\bibinfo {title} {{Thermal {H}all Effect, Spin {N}ernst Effect, and
  Spin Density Induced by a Thermal Gradient in Collinear Ferrimagnets from
  Magnon--Phonon Interaction}},\ }\href
  {https://doi.org/10.1021/acs.nanolett.0c00363} {\bibfield  {journal}
  {\bibinfo  {journal} {Nano Lett.}\ }\textbf {\bibinfo {volume} {20}},\
  \bibinfo {pages} {2741} (\bibinfo {year} {2020})}\BibitemShut {NoStop}%
\bibitem [{\citenamefont {Kl\o{}getvedt}\ and\ \citenamefont
  {Qaiumzadeh}(2023)}]{Klogetvedt2023}%
  \BibitemOpen
  \bibfield  {author} {\bibinfo {author} {\bibfnamefont {J.~N.}\ \bibnamefont
  {Kl\o{}getvedt}}\ and\ \bibinfo {author} {\bibfnamefont {A.}~\bibnamefont
  {Qaiumzadeh}},\ }\bibfield  {title} {\bibinfo {title} {{Tunable topological
  magnon-polaron states and intrinsic anomalous {H}all phenomena in
  two-dimensional ferromagnetic insulators}},\ }\href
  {https://doi.org/10.1103/PhysRevB.108.224424} {\bibfield  {journal} {\bibinfo
   {journal} {Phys. Rev. B}\ }\textbf {\bibinfo {volume} {108}},\ \bibinfo
  {pages} {224424} (\bibinfo {year} {2023})}\BibitemShut {NoStop}%
\bibitem [{\citenamefont {Bao}\ \emph {et~al.}(2023)\citenamefont {Bao},
  \citenamefont {Gu}, \citenamefont {Shangguan}, \citenamefont {Huang},
  \citenamefont {Liao}, \citenamefont {Zhao}, \citenamefont {Zhang},
  \citenamefont {Dong}, \citenamefont {Wang}, \citenamefont {Kajimoto},
  \citenamefont {Nakamura}, \citenamefont {Fennell}, \citenamefont {Yu},
  \citenamefont {Li},\ and\ \citenamefont {Wen}}]{Bao2023}%
  \BibitemOpen
  \bibfield  {author} {\bibinfo {author} {\bibfnamefont {S.}~\bibnamefont
  {Bao}}, \bibinfo {author} {\bibfnamefont {Z.-L.}\ \bibnamefont {Gu}},
  \bibinfo {author} {\bibfnamefont {Y.}~\bibnamefont {Shangguan}}, \bibinfo
  {author} {\bibfnamefont {Z.}~\bibnamefont {Huang}}, \bibinfo {author}
  {\bibfnamefont {J.}~\bibnamefont {Liao}}, \bibinfo {author} {\bibfnamefont
  {X.}~\bibnamefont {Zhao}}, \bibinfo {author} {\bibfnamefont {B.}~\bibnamefont
  {Zhang}}, \bibinfo {author} {\bibfnamefont {Z.-Y.}\ \bibnamefont {Dong}},
  \bibinfo {author} {\bibfnamefont {W.}~\bibnamefont {Wang}}, \bibinfo {author}
  {\bibfnamefont {R.}~\bibnamefont {Kajimoto}}, \bibinfo {author}
  {\bibfnamefont {M.}~\bibnamefont {Nakamura}}, \bibinfo {author}
  {\bibfnamefont {T.}~\bibnamefont {Fennell}}, \bibinfo {author} {\bibfnamefont
  {S.-L.}\ \bibnamefont {Yu}}, \bibinfo {author} {\bibfnamefont {J.-X.}\
  \bibnamefont {Li}},\ and\ \bibinfo {author} {\bibfnamefont {J.}~\bibnamefont
  {Wen}},\ }\bibfield  {title} {\bibinfo {title} {{Direct observation of
  topological magnon polarons in a multiferroic material}},\ }\href
  {https://doi.org/10.1038/s41467-023-41791-9} {\bibfield  {journal} {\bibinfo
  {journal} {Nature Commun.}\ }\textbf {\bibinfo {volume} {14}},\ \bibinfo
  {pages} {6093} (\bibinfo {year} {2023})}\BibitemShut {NoStop}%
\bibitem [{\citenamefont {Mook}\ \emph {et~al.}(2019)\citenamefont {Mook},
  \citenamefont {Henk},\ and\ \citenamefont {Mertig}}]{Mook2019}%
  \BibitemOpen
  \bibfield  {author} {\bibinfo {author} {\bibfnamefont {A.}~\bibnamefont
  {Mook}}, \bibinfo {author} {\bibfnamefont {J.}~\bibnamefont {Henk}},\ and\
  \bibinfo {author} {\bibfnamefont {I.}~\bibnamefont {Mertig}},\ }\bibfield
  {title} {\bibinfo {title} {{Thermal {H}all effect in noncollinear coplanar
  insulating antiferromagnets}},\ }\href
  {https://doi.org/10.1103/PhysRevB.99.014427} {\bibfield  {journal} {\bibinfo
  {journal} {Phys. Rev. B}\ }\textbf {\bibinfo {volume} {99}},\ \bibinfo
  {pages} {014427} (\bibinfo {year} {2019})}\BibitemShut {NoStop}%
\bibitem [{\citenamefont {Chang}\ \emph {et~al.}(2023)\citenamefont {Chang},
  \citenamefont {Liu},\ and\ \citenamefont {MacDonald}}]{Chang2023}%
  \BibitemOpen
  \bibfield  {author} {\bibinfo {author} {\bibfnamefont {C.-Z.}\ \bibnamefont
  {Chang}}, \bibinfo {author} {\bibfnamefont {C.-X.}\ \bibnamefont {Liu}},\
  and\ \bibinfo {author} {\bibfnamefont {A.~H.}\ \bibnamefont {MacDonald}},\
  }\bibfield  {title} {\bibinfo {title} {Colloquium: Quantum anomalous {H}all
  effect},\ }\href {https://doi.org/10.1103/RevModPhys.95.011002} {\bibfield
  {journal} {\bibinfo  {journal} {Rev. Mod. Phys.}\ }\textbf {\bibinfo {volume}
  {95}},\ \bibinfo {pages} {011002} (\bibinfo {year} {2023})}\BibitemShut
  {NoStop}%
\bibitem [{\citenamefont {Zhang}(2016)}]{Zhang2016}%
  \BibitemOpen
  \bibfield  {author} {\bibinfo {author} {\bibfnamefont {L.}~\bibnamefont
  {Zhang}},\ }\bibfield  {title} {\bibinfo {title} {{Berry curvature and
  various thermal Hall effects}},\ }\href
  {https://doi.org/10.1088/1367-2630/18/10/103039} {\bibfield  {journal}
  {\bibinfo  {journal} {New J. Phys.}\ }\textbf {\bibinfo {volume} {18}},\
  \bibinfo {pages} {103039} (\bibinfo {year} {2016})}\BibitemShut {NoStop}%
\bibitem [{\citenamefont {Matsumoto}\ \emph {et~al.}(2014)\citenamefont
  {Matsumoto}, \citenamefont {Shindou},\ and\ \citenamefont
  {Murakami}}]{Matsumoto2014}%
  \BibitemOpen
  \bibfield  {author} {\bibinfo {author} {\bibfnamefont {R.}~\bibnamefont
  {Matsumoto}}, \bibinfo {author} {\bibfnamefont {R.}~\bibnamefont {Shindou}},\
  and\ \bibinfo {author} {\bibfnamefont {S.}~\bibnamefont {Murakami}},\
  }\bibfield  {title} {\bibinfo {title} {{Thermal Hall effect of magnons in
  magnets with dipolar interaction}},\ }\href
  {https://doi.org/10.1103/PhysRevB.89.054420} {\bibfield  {journal} {\bibinfo
  {journal} {Phys. Rev. B}\ }\textbf {\bibinfo {volume} {89}},\ \bibinfo
  {pages} {054420} (\bibinfo {year} {2014})}\BibitemShut {NoStop}%
\bibitem [{\citenamefont {Murakami}\ and\ \citenamefont
  {Okamoto}(2017)}]{Murakami2017}%
  \BibitemOpen
  \bibfield  {author} {\bibinfo {author} {\bibfnamefont {S.}~\bibnamefont
  {Murakami}}\ and\ \bibinfo {author} {\bibfnamefont {A.}~\bibnamefont
  {Okamoto}},\ }\bibfield  {title} {\bibinfo {title} {Thermal {H}all effect of
  magnons},\ }\href {https://doi.org/10.7566/JPSJ.86.011010} {\bibfield
  {journal} {\bibinfo  {journal} {J. Phys. Soc. Jpn.}\ }\textbf {\bibinfo
  {volume} {86}},\ \bibinfo {pages} {011010} (\bibinfo {year}
  {2017})}\BibitemShut {NoStop}%
\bibitem [{\citenamefont {Zhang}\ \emph {et~al.}(2024)\citenamefont {Zhang},
  \citenamefont {Gao},\ and\ \citenamefont {Chen}}]{Zhang2024}%
  \BibitemOpen
  \bibfield  {author} {\bibinfo {author} {\bibfnamefont {X.-T.}\ \bibnamefont
  {Zhang}}, \bibinfo {author} {\bibfnamefont {Y.~H.}\ \bibnamefont {Gao}},\
  and\ \bibinfo {author} {\bibfnamefont {G.}~\bibnamefont {Chen}},\ }\bibfield
  {title} {\bibinfo {title} {{Thermal Hall effects in quantum magnets}},\
  }\href {https://doi.org/https://doi.org/10.1016/j.physrep.2024.03.004}
  {\bibfield  {journal} {\bibinfo  {journal} {Phys. Rep.}\ }\textbf {\bibinfo
  {volume} {1070}},\ \bibinfo {pages} {1} (\bibinfo {year} {2024})}\BibitemShut
  {NoStop}%
\bibitem [{\citenamefont {Go}\ and\ \citenamefont {Kim}(2022)}]{Go2022}%
  \BibitemOpen
  \bibfield  {author} {\bibinfo {author} {\bibfnamefont {G.}~\bibnamefont
  {Go}}\ and\ \bibinfo {author} {\bibfnamefont {S.~K.}\ \bibnamefont {Kim}},\
  }\bibfield  {title} {\bibinfo {title} {{Tunable large spin {N}ernst effect in
  a two-dimensional magnetic bilayer}},\ }\href
  {https://doi.org/10.1103/PhysRevB.106.125103} {\bibfield  {journal} {\bibinfo
   {journal} {Phys. Rev. B}\ }\textbf {\bibinfo {volume} {106}},\ \bibinfo
  {pages} {125103} (\bibinfo {year} {2022})}\BibitemShut {NoStop}%
\bibitem [{\citenamefont {Li}\ \emph {et~al.}(2020)\citenamefont {Li},
  \citenamefont {Sandhoefner},\ and\ \citenamefont {Kovalev}}]{Li2020}%
  \BibitemOpen
  \bibfield  {author} {\bibinfo {author} {\bibfnamefont {B.}~\bibnamefont
  {Li}}, \bibinfo {author} {\bibfnamefont {S.}~\bibnamefont {Sandhoefner}},\
  and\ \bibinfo {author} {\bibfnamefont {A.~A.}\ \bibnamefont {Kovalev}},\
  }\bibfield  {title} {\bibinfo {title} {{Intrinsic spin {N}ernst effect of
  magnons in a noncollinear antiferromagnet}},\ }\href
  {https://doi.org/10.1103/PhysRevResearch.2.013079} {\bibfield  {journal}
  {\bibinfo  {journal} {Phys. Rev. Res.}\ }\textbf {\bibinfo {volume} {2}},\
  \bibinfo {pages} {013079} (\bibinfo {year} {2020})}\BibitemShut {NoStop}%
\end{thebibliography}
\end{document}